\documentclass[useAMS, usenatbib]{mn2e}

\usepackage[dvips]{graphicx,color}
\usepackage[]{txfonts}

\def\simless{\mathbin{\lower 3pt\hbox
{$\rlap{\raise 5pt\hbox{$\char'074$}}\mathchar"7218$}}}   
\def\simmore{\mathbin{\lower 3pt\hbox
{$\rlap{\raise 5pt\hbox{$\char'076$}}\mathchar"7218$}}}   

\def\difd{\mathrm{d}}
\newcommand{\dsfrac}[2]{\displaystyle{\frac{#1}{#2}}}
\newcommand{\eref}[1]{(\ref{#1})}

\usepackage{enumerate}

\def\xcd{x'}
\def\tcd{t'}
\def\dop{{\cal D}}

\newcommand{\corr}[1]{{#1}}

\title[Radiative signatures of magnetic fields]{Radiative signature of magnetic fields in internal shocks}
\author[P. Mimica and M. A. Aloy]{P. Mimica$^{1}$\thanks{E-mail: Petar.Mimica@uv.es} and M. A. Aloy$^{1}$\\
$^{1}$Departamento de Astronom\'ia y Astrof\'isica, Universidad de Valencia, 46100, Burjassot, Spain}

\begin{document}

\maketitle

\label{firstpage}

\begin{abstract}
  Common models of blazars and gamma-ray bursts assume that the plasma
  underlying the observed phenomenology is magnetized to some
  extent. Within this context, radiative signatures of dissipation of
  kinetic and \corr{conversion of} magnetic energy in internal shocks
  of relativistic magnetized outflows are studied. We model internal
  shocks as being caused by collisions of homogeneous plasma
  shells. We compute the flow state after the shell interaction by
  solving Riemann problems at the contact surface between the
  colliding shells, and then compute the emission from the resulting
  shocks. Under the assumption of a constant flow luminosity we find
  that there is a clear difference between the models where both
  shells are weakly magnetized ($\sigma\simless 10^{-2}$) and those
  where, at least, one shell has a $\sigma\simmore 10^{-2}$. We obtain
  that the radiative efficiency is largest for models in which,
  regardless of the ordering, one shell is weakly and the other
  strongly magnetized. Substantial differences between weakly and
  strongly magnetized shell collisions are observed in the
  inverse-Compton part of the spectrum, as well as in the optical,
  X-ray and $1$GeV light curves. We propose a way to distinguish
  observationally between weakly magnetized from magnetized internal
  shocks by comparing the maximum frequency of the inverse-Compton and
  synchrotron part of the spectrum to the ratio of the inverse-Compton
  and synchrotron fluence. Finally, our results suggest that LBL
  blazars may correspond to barely magnetized flows, while HBL blazars
  could correspond to moderately magnetized ones. Indeed, by comparing
  with actual blazar observations we conclude that the magnetization
  of typical blazars is $\sigma\simless 0.01$ \corr{for the internal
    shock model to be valid in these sources}.
\end{abstract}


\section{Introduction}
\label{sec:introduction}

Highly variable radiation flux has been observed in the relativistic
outflows of blazars and gamma-ray bursts (GRBs). Even though the
radiation energy and time scales are different for both classes of
objects ($\gamma$-rays on a millisecond timescale for GRBs versus
X-rays on a timescale of hours for blazars) the underlying physics
responsible for the energy dissipation might be very similar. The
internal shock scenario \citep{Rees:1994ca} has been used to explain
the variability of blazars \citep[e.g.,][]{Spada:2001do,Mimica:2004ay}
and GRBs
\citep[e.g.,][]{Kobayashi:1997vf,Daigne:1998wq,Bosnjak:2009dv}. In
this scenario inhomogeneities in a relativistic outflow cause parts of
the fluid to collide and produce shocks waves which dissipate
energy. The shell collisions are often idealized as collisions of dense
shells. In recent years one- and two-dimensional relativistic
hydrodynamics \citep[RHD,][]{Kino:2004in,Mimica:2004ay,Mimica:2005aa}
and relativistic magneto-hydrodynamics \citep[RMHD,][]{Mimica:2007aa}
simulations of the shell collisions have been performed and have
showed that the dynamics of shell interaction is much more complex
than what is commonly assumed when modeling shell interactions
analytically
\citep[e.g.,][]{Kobayashi:1997vf,Daigne:1998wq,Spada:2001do,
  Bosnjak:2009dv}. Particularly, the influence of the magnetic field
(if present) has been shown to significantly alter the dynamics
\citep{Mimica:2007aa}. In spite of these efforts, we still do not
known with certainty whether the flow, whose energy is being
dissipated, is significantly magnetized, or whether it is only the
kinetic energy which ultimately powers the emission. 

In a previous work \citep[][MA10 in the following]{Mimica:2010aa} we
have studied the dynamic efficiency, i.e. the efficiency of conversion
of kinetic to thermal and/or magnetic energy in internal shocks. We
found that the dynamic efficiency is actually higher if the shells are
moderately magnetized ($\sigma \approx 0.1$, see the next section for
the definition of $\sigma$) than if both are unmagnetized. However, we
did not compute the radiative efficiency of such interactions, but
instead used the dynamic efficiency as an upper bound of it. Recently
\citet[][BD10 in the following]{Bottcher:2010gn}, \citet{Joshi:2011bp}
and \citet{Chen:2011} have presented sophisticated models for the
detailed computation of the emission from internal shocks.  While
these models assume a simple hydrodynamic evolution, they employ a
time-dependent radiative transfer scheme which involves the
synchrotron and synchrotron self-Compton (SSC) processes as well as
the contribution Comptonized external radiation (external inverse
Compton - EIC), all the while taking into account the radiative losses
of the emitting non-thermal particles. We have adapted the method of
BD10 and use it to perform a parametric study, addressed to infer the
magnetization of the flow from the light curves and spectra of
internal shocks in magnetized plasma.

The organization of this paper is as follows:
Section~\ref{sec:dynamics} briefly summarizes the model of MA10 which
is used to study the shell collision dynamics, and in
Sections~\ref{sec:particles} and \ref{sec:radiation} we describe the
numerical method we employ to compute the non-thermal radiation. We
discuss the radiative efficiency in Section~\ref{sec:efficiency} and
present the spectra and light curves in Section~\ref{sec:emission}. A
global parameter study is elaborated in Section~\ref{sec:params}. We
close the paper with a discussion of our results and give our
conclusions (Sec.~\ref{sec:discussion}).

\section{Shell collision dynamics}
\label{sec:dynamics}

As was discussed in detail in the Section~2 of MA10, our aim is to
model a large number of shell collisions with varying
properties. Therefore, we employ a simplified model for a single shell
collision, based on the exact solution of the Riemann problem. When
describing the initial states of the Riemann problem we will use
subscripts L and R to denote left (faster) and right (slower) shells,
respectively.

We assume a cylindrical outflow with a radius
$R$. \corr{\citet{Mimica:2004ay} show that the jet lateral expansion
  in this case is negligible. For simplicity, and being consistent
  with previous work in the field (e.g., BD10,
  \citealt{Joshi:2011bp}), we also ignore the shell longitudinal
  expansion after the shocks cross the shell (see also
  Section~\ref{sec:evolution}).} Following the equation 9 of MA10 we
define the luminosity as
\begin{equation}\label{eq:luminosity}
L:=\pi R^2 \rho c^3 \left[\Gamma^2(1+\epsilon+\chi+\sigma)-\Gamma
\right] \sqrt{1 - \Gamma^{-2}}\ ,
\end{equation}
where $c$ is the speed of light in vacuum, $\rho$ is the fluid
rest-mass density, $\epsilon$ is the specific internal energy,
$\chi:=p/(\rho c^2)$ is the initial ratio between the thermal pressure
and the rest-mass energy density, and $\sigma:=B^2 / (4\pi\rho \Gamma^2
c^2)$ is the magnetization parameter. Here $B$ is the strength of the
large-scale magnetic field, which is perpendicular to the direction of
propagation of the fluid moving with velocity $v$ and a corresponding
Lorentz factor $\Gamma:=1 / \sqrt{1 - (v/c)^2}$. The specific internal
energy is related to the pressure and to the density through the
equation of state. We use the TM analytic approximation to the Synge
equation of state \citep{deBerredoPeixoto:2005if,Mignone:2005hc} and
obtain:
\begin{equation}\label{eq:epsilon}
\epsilon:=\dsfrac{3}{2}\dsfrac{p}{\rho c^2} +
\left[\dsfrac{9}{4}\left(\dsfrac{p}{\rho c^2}\right)^2 +
  1\right]^{1/2} - 1\ .
\end{equation}
We assume that $L_L=L_R$ and $\chi_L=\chi_R$. Furthermore, as in MA10,
we assume $\Gamma_L :=(1+\Delta g)\Gamma_R$. This leaves us with $R$,
$\sigma_L$, $\sigma_R$ and $\Delta g$ as parameters, because all other
quantities can be determined using equations \eref{eq:luminosity} and
\eref{eq:epsilon}. To these, we add an additional parameter $\Delta
r$, which is the initial width of the shells in the LAB-frame. While
it does not influence the solution of the Riemann problem, it provides
the physical scale necessary for the calculation of the observed
emission.

Once the initial states are constructed, we compute the exact solution
of the Riemann problem using the solver of \citet{Romero:2005zr}. The
initial discontinuity between left and right states decomposes into a
contact discontinuity (CD), and a left-going and a right wave (in the
frame in which the CD is at rest). Depending on the particulars of the
initial states these waves can either be shocks or rarefactions. We
label the left-going wave with RS to denote a reverse shock, and with
RR in case a reverse rarefaction happens. Similarly, we label the
right-going wave with FS or FR to differentiate the cases in which a
forward shock or a forward rarefaction occurs, respectively. We will
use a subscript $S$ to refer to the properties of the shocked
fluid in general, and the subscripts $FS$ and $RS$ when distinguishing
between the front and reverse shocked fluids. Finally, we will use
the subscript $0$ for properties of the initial states in general, and
the subscripts $L$ and $R$ when we need to distinguish between left
and right initial states. Because we assume that the flow luminosity
is the same for both initial states, using \eref{eq:luminosity} we
determine the number density in the shells to be
\begin{equation}\label{eq:numdens}
  n_{L,R} = \dsfrac{L}{\pi R^2 m_p c^3
    \left[\Gamma_{L,R}^2(1+\epsilon+\chi+\sigma_{L,R})-\Gamma_{L,R}
    \right] \sqrt{1 - \Gamma_{L,R}^{-2}}}\ ,
\end{equation}
where $\Gamma_L = (1+\Delta g)\Gamma_R$.

The Riemann solver provides us with the bulk velocity of the shocked
fluid $\beta c$ (and its Lorentz factor $\Gamma =(1-\beta^2)^{-1/2}$),
and velocities $\beta_{FS}c$ and $\beta_{RS} c$ of the FS and RS,
respectively (provided they exist).  The velocity of the initial
(unshocked) states in the CD rest frame is
\begin{equation}\label{eq:beta0}
  \beta'_0=\dsfrac{\beta_0 - \beta}{1 - \beta\beta_0}\ .
\end{equation}
The shock velocities in the frame of the CD can be computed as
\begin{equation}\label{eq:betas}
  \beta'_S = \dsfrac{\beta_S - \beta}{1 - \beta\beta_S}  \ ,
\end{equation}
where prime denotes quantities in the CD rest frame. In this frame
the shock crosses the shell at a time
\begin{equation}\label{eq:tcross}
  t'_{\rm cross, S}=\dsfrac{\Delta
      r'_0}{c\left|\beta'_S\right|} \ ,
\end{equation}
where $\Delta r'_0$ is the shell width in the CD frame,
\begin{equation}\label{eq:deltar}
  \Delta r'_0 =\Gamma \Delta r \dsfrac{\beta -
    \beta_S}{\beta_0 - \beta_S}\ .
\end{equation}

\section{Non-thermal particles}
\label{sec:particles}

In this section we show the properties of non-thermal particles and
their emission. We first discuss the model for the magnetic field and
non-thermal particles, and then outline the method used to compute
their emission.

\subsection{Magnetic field}
\label{sec:field}

As in \cite{Mimica:2010aa} and BD10, we assume that there exists a
stochastic magnetic field, which is created {\it in situ} by the
shocks arising in the collision of the shells. We label this field by
$B_{\rm S, st}$, and by definition its strength is a fraction
$\epsilon_B$ of the internal energy density of the shocked shell
$u_S$ (obtained, in our case, by the exact Riemann solver):
\begin{equation}\label{eq:Bmic}
  B_{\rm S, st} = \sqrt{8\pi \epsilon_B u_S}\, .
\end{equation}

Since we allow for arbitrarily magnetized shells, there is also an
ordered (macroscopic) magnetic field component $B_{\rm S, mac}$, which
is a direct output of the exact Riemann solver. The total magnetic
field is then $B_S:= \sqrt{B_{\rm S, st}^2 + B_{\rm S, mac}^2}$.

\subsection{Injection spectrum of non-thermal particles}
\label{sec:injection}

We assume that a fraction of electrons in the unshocked shell is
accelerated to high energies at the shock front. Following Section 3
of BD10, we assume that a fraction $\epsilon_e$ of the dissipated
kinetic energy is used to accelerate electrons. We assume that some
particle acceleration mechanism operates at shocks, such that a
fraction of the electrons in the unshocked shell is accelerated to
high energies in the vicinity of the shock front. As it is commonly
done \citep[][BD10]{Bykov:1996aa,Daigne:1998wq,Mimica:2004ay}, we
assume that a fraction $\epsilon_e$ of the dissipated kinetic energy
is used to accelerate electrons.
 The width of the
acceleration zone $\Delta r'_{\rm acc}$ is parametrized as a multiple
$\Delta_{\rm acc}$ of the proton Larmor radius in the shocked
fluid. Therefore, the acceleration at a given point in the shocked
fluid lasts for a time $\Delta t'_{\rm acc} = \Delta r'_{\rm acc} /
(\beta'_S c)$. We have
\begin{equation}\label{eq:deltaracc}
  \Delta r'_{\rm acc} = \Delta_{\rm acc}\dsfrac{\Gamma'_0 m_p
    c^2}{e B_S}\ ,
\end{equation}
From this expression, we compute the volume where acceleration takes place as $V'_{\rm acc} = \pi R^2
\Delta r'_{\rm acc}$. The energy injection rate into the acceleration
region is
\begin{equation}\label{eq:einj}
  \dsfrac{\mathrm{d}E'_{\rm inj,0}}{\mathrm{d}t'} = \pi R^2 \epsilon_e
  u_S \dsfrac{\Delta r'_{\rm acc}}{\Delta t'_{\rm acc}}\, ,
\end{equation}
and we assume that the energy spectrum of the injected relativistic
particles is a power-law in the electron Lorentz factor $\gamma$,
\begin{equation}\label{eq:elinj}
\dsfrac{\mathrm{d}n'_{\rm inj}}{\mathrm{d}t'\ \mathrm{d}\gamma} = Q_0
\gamma^{-q} H(\gamma;\ \gamma_1, \gamma_2)\ ,
\end{equation}
where $n'_{\rm inj}$ is the number density of the injected electrons, $Q_0$
is a normalization factor and $\gamma_1$ and $\gamma_2$ are the
lower and upper injection cutoffs (computed below), all measured in
the shocked fluid rest frame. The step function is defined as usual by
$H(x; a, b) = 1$ if $a\leq x\leq b$ and
$0$ otherwise.

\corr{A cautionary note should be added here regarding the fact that
  we choose that the spectral energy distribution of the injected
  particles is a pure power-law even in the high-$\sigma$ regime. Both
  theoretical arguments \citep[e.g.,][]{Kirk:1989yt} and recent
  particle-in-cell (PIC) simulations \citep[e.g.,][]{Sironi:2009aa}
  have shown that particle acceleration is not very efficient in the
  presence of a strong magnetic field parallel to the shock front. The
  modifications to the particle injection spectrum might involve the
  presence of the thermal population. Recently, a calculation by
  \citet{Giannios:2009zm} shows how the spectrum of the GRB prompt
  emission might look in such a case: a bump at the spectral maximum
  and a lower contribution at ultra-high energies. However, current
  PIC simulations have not been run for sufficiently long time to
  achieve a stable situation. Thus, the fraction of the energy which
  goes into thermal electrons (parameter $\delta$ of
  \citealt{Giannios:2009zm}) is still to be determined. In this
  particular study we set this fraction to zero and thus avoid
  introducing another free parameter in our models.}

Integrating the distribution Eq.~\ref{eq:elinj} in Lorentz factor and
equating the result to Eq.~\ref{eq:einj} divided by $V'_{\rm acc}$ (in
order to obtain the energy density injection rate into the
acceleration region) we can compute the normalization factor for the
electron injection,
\begin{equation}\label{eq:Q0}
  Q_0 = \dsfrac{\mathrm{d}E'_{\rm
        inj,0}/\mathrm{d}t'}{V'_{\rm acc}m_e c^2}\times \left\{
      \begin{array}{rl}
        \dsfrac{q - 2}{\gamma_1^{2-q} - \gamma_2^{2-q}} & \mathrm{if}\
        q\neq 2 \\[4mm]
        1/\ln\left(\dsfrac{\gamma_2}{\gamma_1}\right) & \mathrm{if}\ q=2
        \end{array}
        \right. \ .
\end{equation}

\subsection{Particle injection cut-offs}
\label{sec:cutoffs}

As was done in \cite{Mimica:2010cc}, we assume that the upper cutoff for
the electron injection is obtained by assuming that the acceleration
time scale is proportional to the gyration time scale. Then the
maximum Lorentz factor is obtained by equating this time scale to the
cooling time scale,
\begin{equation}\label{eq:gamma2}
  \gamma_2 = \left(\dsfrac{3m_e^2 c^4}{4\pi a_{\rm acc} e^3 B_S}\right)^{1/2}\, .
\end{equation}
where $a_{\rm acc}\geq 1$ is the acceleration efficiency parameter
(BD10). The lower cut-off is obtained by assuming, in complete analogy
to Eq.~\ref{eq:einj}, that the number of accelerated electrons is
related to the number of electrons passing through the shock front,
\begin{equation}\label{eq:ninj}
    \dsfrac{\mathrm{d}N'_{\rm inj,0}}{\mathrm{d}t'} = \zeta_e \pi
  R^2 n_0 \Gamma'_0 \beta'_S c\ ,
\end{equation}
where $\zeta_e$ is the fraction of electrons accelerated into the
power-law distribution. From Eqs.~\ref{eq:ninj}, \ref{eq:einj} and
\ref{eq:elinj} we get
\begin{equation}\label{eq:gamma1}
\dsfrac{\int_{\gamma_1}^{\gamma_2}\ \mathrm{d}\gamma\
  \gamma^{1-q}}{\int_{\gamma_1}^{\gamma_2}\ \mathrm{d}\gamma\
  \gamma^{-q}} =
\dsfrac{\epsilon_e}{\zeta_e}\dsfrac{u_S}{\Gamma'_0n_0 m_e
  c^2}\, .
\end{equation}
Since we are dealing with potentially highly magnetized fluids, the
condition $\gamma_2\gg \gamma_1$ cannot be assumed (see
Eq.~\ref{eq:gamma2}), and therefore we cannot use the equation such as
Eq.~13 of BD10. Therefore we compute $\gamma_1$ from
Eq.~\ref{eq:gamma1} numerically using an iterative procedure.

\subsection{Evolution of the particle distribution}
\label{sec:evolution}

In this work we assume that particles cool via synchrotron and
external-Compton processes. \corr{We ignore the adiabatic cooling in
  this work since we are primarily interested in collisions of
  magnetized shells, where the electrons are fast-cooling. The
  consequence of not accounting for the adiabatic loses of the
  particle distribution is that our method overestimates the emission
  after the shocks cross the shells. Nevertheless, most of the
  features that make substantive differences between the dynamics
  triggered by magnetized and non-magnetized shells happen in the
  early light curve and, thus, neglecting the expansion of the shells
  plasma does not change the qualitative conclusions of this paper.}

The radiative losses for an electron with a Lorentz factor $\gamma$
can be written as
\begin{equation}\label{eq:radloss}
  \dot{\gamma} = -\dsfrac{4}{3}c\sigma_T\dsfrac{u'_B + u'_{\rm ext}}{m_e c^2}\gamma^2\, ,
\end{equation}
where $u'_B = B_S^2 / 8\pi\Gamma^2$ and $u'_{\rm ext}$ are the energy
density of the magnetic field and the external radiation field (see
Sec.~\ref{sec:EIC}) in the shocked fluid frame, respectively. Once the
energy losses have been specified, we use the semi-analytic solver of
\citet{Mimica:2004ay} to compute the particle distribution at any time
after the start of the injection at shock. More precisely, we use the
solution for the continuous injection and particle cooling
\citep[Eq.~19 of][]{Mimica:2004ay} for the time $\Delta t'_{\rm acc}$
since the beginning of the shock acceleration. After that time the
shock acceleration ends and we approximate the resulting particle
distribution by a piecewise power-law function. Then we employ Eq.~17
of \citet{Mimica:2004ay} on each of the power-law segments to compute
the subsequent evolution.

\section{Non-thermal radiation}
\label{sec:radiation}

We assume that the observer's line of sight makes an angle $\theta$
with the jet axis, which is also the direction of propagation of the
fluid. We denote by $x$ and $t$ the position and time in the observer
frame, and by $\xcd$ and $\tcd$ the location and time in the CD
frame. We assume that the CD is located at $\xcd=0$ for all
$\tcd$. Let $\mu$ be $\mu:=\cos\theta$, and define the time at which an observer sees
the radiation emitted from $x$ at time $t$ as 
\begin{equation}\label{eq:tobs}
T = t - x\mu/c\, .
\end{equation}
Then
the time as a function of the time of observation and position can be
written as
\begin{equation}\label{eq:tcd}
   \tcd = \dop \left(T / (1 + z) + \Gamma(\mu - \beta) \xcd / c \right)\, ,
\end{equation}
where $\dop := [\Gamma(1 - \beta\mu)]^{-1}$ is the Doppler factor and
$z$ is the redshift. Lorentz transformations have been applied to
Eq.~\ref{eq:tobs} to obtain Eq.~\ref{eq:tcd}.

An important quantity is the time elapsed after the shock has passed a
given $\xcd$. From such value one can calculate the \emph{age} of the
electron distribution function at that position, which turns to be
the time since the shock acceleration has begun. Thus, the age can be
defined as
\begin{equation}\label{eq:agedef}
  \tcd_a := \tcd - \dsfrac{\xcd}{\beta'_S c} \, .
\end{equation}
A more useful expression involves $T$. Using Eqs.~\ref{eq:agedef} and
\ref{eq:tcd} we get, for the FS
\begin{equation}\label{eq:ageFS}
  \tcd_{a,FS} = \dop \left[\dsfrac{T}{1 + z} - \dsfrac{\xcd}{c\Gamma}\dsfrac{1 -
      \beta_{FS}\mu}{\beta_{FS} - \beta}\right]\, ,
\end{equation}
and for the RS
\begin{equation}\label{eq:ageRS}
  \tcd_{a,RS} = \dop \left[\dsfrac{T}{1 + z} - \dsfrac{\xcd}{c\Gamma}\dsfrac{1 -
      \beta_{RS}\mu}{\beta_{RS} - \beta}\right]\, .
\end{equation}
We note that Eq.~\ref{eq:ageFS} has to be used when $\xcd\geq 0$,
while Eq.~\ref{eq:ageRS} is valid when $\xcd<0$. If $\tcd_a\leq 0$,
then the shock has not crossed that position yet and, consequently,
that place does not contribute to the emission yet.

The observed luminosity in the CD rest frame is
\begin{equation}\label{eq:lumCD}
  \nu' L'_{\nu'}(T) = \pi R^2 \int\limits_{\xcd_{\rm
      min}(T)}^{\xcd_{\rm max}(T)}\difd \xcd\ \nu'
  j'_{\nu'}[\tcd_a(T, \xcd)]\ ,
\end{equation}
where the lower and upper limits depend on (1) whether the shock
exists, and (2) whether it has crossed the shell\footnote{\corr{Since
    the assumed geometry in this model is cylindrical, the effects of
    the high-latitude emission are ignored. This means that all the
    peaks and breaks in the light curves are sharper than would be in
    case a conical geometry was assumed. Another consequence of the
    assumed geometry is that we overestimate the rate at which the
    light curve declines after shocks cross the shells (see
    Section~\ref{sec:lcs}).}}. If the RS does not exist, then
$\xcd_{\rm min} = 0$; otherwise it is
\begin{equation}\label{eq:xmin}
  \xcd_{\rm min}(T) = \max\left(\dsfrac{\Gamma cT}{1 + z}\dsfrac{\beta_{RS} - \beta}{1 -
        \beta_{RS}\mu}, -\Delta r'_L\right)\, ,
\end{equation}
where $\Delta r'_L$ is the faster
shell width in the CD frame. Analogously, for $\xcd_{\rm max} = 0$ the
FS is non-existent; otherwise it is
\begin{equation}\label{eq:xmax}
  \xcd_{\rm max}(T) = \min\left(\dsfrac{\Gamma cT}{1 + z}\dsfrac{\beta_{FS} - \beta}{1 -
        \beta_{FS}\mu}, \Delta r'_R\right)\, ,
\end{equation}
$\Delta r'_R$ being the slower shell width in the CD frame. We point
out that, to perform the integral in Eq.~\ref{eq:lumCD}, $
j'_{\nu'}(t'_a)$ should be computed for the particle distribution
evolved using values for the FS if $\xcd>0$, and RS if $\xcd<0$
(i.e. in Eqs.~\ref{eq:Q0}, \ref{eq:gamma2}, \ref{eq:gamma1} and
\ref{eq:Bmic} the values for the corresponding shocked fluid should be
used).

Considering that $\nu' = \nu (1 + z) / \dop$, and using
Eq.~\ref{eq:lumCD}, we can compute the flux in the observer frame
obtaining \citep[BD10, ][]{Dermer:2008ik}
\begin{equation}\label{eq:fluxobs}
  \nu F_\nu(T) = \dsfrac{\dop^4 \pi R^2}{d_L^2} \int\limits_{\xcd_{\rm
      min}(T)}^{\xcd_{\rm max}(T)}\difd \xcd\ \nu'
  j'_{\nu'}[\tcd_a(T, \xcd)]\, ,
\end{equation}
where $d_L$ is the luminosity distance. We perform the integration in
Eq.~\ref{eq:fluxobs} numerically. 

The total emissivity is assumed to be the result of combining three
emission processes: (1) synchrotron radiation, (2) inverse Compton
with an external radiation field (EIC), and the synchrotron
self-Compton (SSC) up-scattering. These emission processes are
considered in more detail in the next sections.

\subsection{Synchrotron emission}
\label{sec:synchrotron}

We compute the synchrotron emission for each power-law segment of the
electron distribution (see Sec.~\ref{sec:evolution}) separately. In
order to speed up the calculation we use the interpolation method
described in \citet[][section 4]{Mimica:2009aa} and, in more detail,
in \citet[][sections 2.1.3 and 4.3.1]{Mimica:2004zz}.

\subsection{EIC emission}
\label{sec:EIC}

Following BD10, we assume that the external radiation field is
monochromatic and isotropic in the AGN frame. We denote the frequency
and the radiation field energy density in this frame by
$\nu_{\rm ext}$ and $u_{\rm ext}$, respectively. Transforming into the
shocked fluid frame we get
\begin{equation}\label{eq:extrad}
  \begin{array}{rcl}
  \nu'_{\rm ext} &=& \Gamma \nu_{\rm ext}\\
  u'_{\rm ext} &=& \Gamma^2 u_{\rm ext}
\end{array}
\end{equation}
Analogously to the computation of the synchrotron emission
(Sec.~\ref{sec:synchrotron}), we compute the EIC emissivity for each
power-law segment separately. We use Eq.~2.94 of
\citet{Mimica:2004zz}, but replacing $I(\nu_0)/\nu_0$ by $c u'_{\rm
  ext} / \nu'_{\rm ext}$ and with an additional cut-off (approximating
the Klein-Nishina decline of the Compton cross-section) such that the
emissivity is zero for $h\nu\geq m_e^2 c^4 / (h \nu'_{\rm ext})$
\citep[see also,][]{Aloy:2008}. \corr{Values of $\nu_{\rm ext}$
  and $u_{\rm ext}$ used in this work can be found in Table 1.}

\subsection{SSC emission}
\label{sec:SSC}

Analogously to Sec.~\ref{sec:EIC}, we use the Eq.~2.94 of
\citet{Mimica:2004zz}. However, in the case of SSC the incoming
intensity of the synchrotron radiation depends on $\xcd$ and $T$. For
a point on the shell axis the (angle averaged) intensity at frequency
$\nu_0$ can be written as
\begin{equation}\label{eq:incint}
  I_{0, \nu_0}(T, \xcd) = \dsfrac{1}{2}\int\limits_{0}^\pi
  \mathrm{d}\theta' \int\limits_0^{L(\theta')}\mathrm{d}s\
  j'_{\nu_0, {\rm syn}}(\tcd(T) - s/c, \xcd + s\cos\theta')\, ,
\end{equation}
where $L(\theta')$ is the length of the segment in direction $\theta'$
from which synchrotron emission has had time to arrive to $\xcd$ at a
time $\tcd$, and $\tcd(T)$ is computed using Eq.~\ref{eq:tcd}. The
synchrotron emissivity $j'_{\nu_0, {\rm syn}}$ can be rewritten
in terms of $T$ using Eqs.~\ref{eq:agedef}, and \ref{eq:ageFS} (or
\ref{eq:ageRS}),
\begin{equation}\label{eq:incemiss}
  \begin{array}{l}
  j'_{\nu_0, {\rm syn}}\left(\tcd(T) - \dsfrac{s}{c}, \xcd +
    s\cos\theta'\right) = \\
  \ \ \ \ j'_{\nu_0, {\rm syn}}\left[t'_a(T, \xcd) -
    \dsfrac{s}{c}\left(1 + \cos\theta' \dsfrac{1 -
        \beta\beta_S}{\beta_S - \beta} \right) \right]\, ,
  \end{array}
\end{equation}
From Eq.~\ref{eq:incemiss} we can see that $L(\theta')$ can be
computed by requiring that the following condition be satisfied for
each $\theta'$,
\[
t'_a(T, \xcd) -
    \dsfrac{s}{c}\left(1 + \cos\theta' \dsfrac{1 -
        \beta\beta_S}{\beta_S - \beta} \right) > 0\, .
\]
If this condition is not satisfied, it means that the shock has not
passed the point $\xcd + s\cos\theta'$ at time $\tcd(T) - s/c$ yet ,
i.e. there is no synchrotron emission from that point to contribute to
the incoming intensity. In addition, we also require that
$L(\theta')\leq R$. Finally, it should not be forgotten that when
$\xcd + s\cos\theta' > 0$ the emissivity of fluid shocked by the FS
should be used, and the one corresponding to the shocked fluid by the
RS otherwise. Also, if either of the shocks is not present, there is
no contribution from the corresponding region.

In practice the numerical cost of a direct evaluation of double
integral in Eq.~\ref{eq:incint} is prohibitive if we take into account
that this intensity has to be evaluated for each $\xcd$ in
Eq.~\ref{eq:fluxobs}. To overcome this problem we approximate
Eq.~\ref{eq:incint} by discretizing the angular integral in a
non-uniform $\theta'$-intervals. The choice of non-uniform intervals
is motivated by the fact that most of the contribution of the incoming
radiation comes from angles close to $\mu = -\beta_S$, so that we
concentrate most of the bins close to that angle. Numerical testing
shows that using $13$ bins provides an acceptable tradeoff between the
accuracy and the computational requirements.

\section{Radiative efficiency}
\label{sec:efficiency}

In this section we compare the radiative efficiency of the internal
shocks with their corresponding dynamic efficiency. We use the
kinematic parameters from MA10 in the blazar regime, while the
parameters used to compute emission are guided by the values from BM10
(see Table~\ref{tab:params} for the complete list).
\begin{table}
  \begin{center}
    \begin{tabular}{|c|c|}
      \hline \hline
      Parameter & value \\
      \hline
      $\Gamma_R$ & $10$ \\
      $\Delta g$ & $1$ \\
      $\sigma_L$ & $[10^{-6}, 10^1]$ \\
      $\sigma_R$ & $[10^{-6}, 10^1]$ \\
      $\epsilon_B$ & $10^{-3}$ \\
      $\epsilon_e$ & $10^{-1}$ \\
      $\zeta_e$ & $10^{-2}$ \\
      $\Delta_{\rm acc}$ & $10$ \\
      $a_{\rm acc}$ & $10^{6}$ \\
      $R$ & $3\times 10^{16}$ cm \\
      $\Delta r$ & $6\times 10^{13}$ cm \\
      $q$ & $2.6$ \\
      $L$ & $5\times 10^{48}$ erg s$^{-1}$ \\
      $u_{\rm ext}$ & $5\times 10^{-4}$ erg cm$^{-3}$ \\
      $\nu_{\rm ext}$ & $10^{14}$ Hz \\
      $z$ & $0.5$ \\
      $\theta$ & $5^o$ \\
      \hline
    \end{tabular}
  \end{center}
  \caption{Blazar model parameters used in this work. Note that $\sigma_L$
    and $\sigma_R$ can vary continuously in the indicated range.}
  \label{tab:params}
\end{table}
All parameters of our models are fixed except for $\sigma_L$ and
$\sigma_R$, which can vary in the range indicated by the
Table~\ref{tab:params}. In the rest of the paper we distinguish models
by the value of the magnetization of each shell, e.g., a model with
$\sigma_L = 0.1$ and $\sigma_R = 1$ is denoted by the pair $(0.1, 1)$.

As can be seen from Eq.~\ref{eq:einj}, in our model only the thermal
energy can be injected into the non-thermal particle population.  We
point out that, alternatively or simultaneously, magnetic dissipation
can provide a source for the emission in internal shocks
\citep[e.g.,][and references therein]{Giannios:2009zi,Nalewajko:2011}, which we are not considering here. Thus, the
radiative efficiency we compute in this paper is only a lower bound to
the actual radiative efficiency of the binary collision of
relativistic magnetized shells. As is shown in the
Appendix~\ref{app:dyneff}, we can use the definition of the dynamic
efficiency inspired by the recent work of \citet{Narayan:2011df}. Its
advantage is the Lorentz-invariance, which enables us to compare it to
the radiative efficiency of our model.

Following MA10, but using the definitions for the different energy
components $\hat{E}_K$, $\hat{E}_T$ and $\hat{E}_M$ of the Appendix
(Eq.~\ref{eq:energiesNKT11}), we denote by $\hat{E_0}$ the total energy in the
unshocked shells,
\begin{equation}\label{eq:E_0}
  \begin{array}{rl}
    \hat{E}_0 &:= \hat{E}_K(\Gamma_R  (1 + \Delta g), n_L m_p, 1) +
    \hat{E}_T(\Gamma_R (1 + \Delta g), n_L m_p, \chi n_L m_p c^2, 1)\\[4mm]
    &+ \hat{E}_M(\Gamma_R (1 + \Delta g), n_L m_p, \sigma_L, 1) +
    \hat{E}_K(\Gamma_R, n_R m_p, 1) \\[4mm]
    &+ \hat{E}_T(\Gamma_R, n_R m_p, \chi n_R m_p c^2, 1) + \hat{E}_M(\Gamma_R, n_R
    m_p, \sigma_R, 1)\, .
    \end{array}\, 
\end{equation}
where $\chi$ is the pressure-to-density ratio of
the cold initial shells and is set to $10^{-4}$. We also define the
width of the shocked shells in terms of their initial width,
\begin{equation}\label{eq:zetaLR}
\zeta_W: =\dsfrac{\left| \beta - \beta_S \right|}{\beta_0 - \beta_S}
\end{equation}
so that $0\leq \zeta_W \leq 1$ \footnote{Note that comparing
  Eqs.~\ref{eq:deltar} and \ref{eq:zetaLR} we
see that $\Delta r_0' = \Gamma \zeta_W \Delta r$, which is just a
Lorentz transformation of the shocked shell width from lab to CD
frame.}. The dynamic thermal efficiency for the faster shell is
defined as
\begin{equation}\label{eq:thermalL}
  \begin{array}{l}
    \epsilon_{T, L} := \dsfrac{1}{\hat{E}_0} \times \\
    \left[\hat{E}_T(\Gamma_{S, L}, n_{S, L} m_p, p_{S,
        L}, \zeta_{W, L}) -
      \hat{E}_T(\Gamma_R (1 + \Delta g), n_L m_p, \chi n_L m_p c^2,
      1)\right]
  \end{array}
\end{equation}
and analogous definitions can be written for $\epsilon_{M, L}$,
$\epsilon_{T, R}$ and $\epsilon_{M, R}$ (see Eqs.~13, 14, 16 and
17 of MA10). The total (Lorentz invariant) dynamic thermal and
magnetic efficiency is
\begin{eqnarray}
  \epsilon_T = \epsilon_{T, L} + \epsilon_{T, R}\, , \\
  \epsilon_M = \epsilon_{M, L} + \epsilon_{M, R}\, .
\end{eqnarray}
From these equations it can be seen that the radiative efficiency can be
at most $\epsilon_e (\epsilon_T + \epsilon_M)$. More formally, we can
write the radiative efficiency as \corr{(neglecting adiabatic cooling)}
\begin{equation}\label{eq:radeff}
  \epsilon_{\rm rad} := \epsilon_e f_{\rm rad} (\epsilon_T + \epsilon_M)\, ,
\end{equation}
where $f_{\rm rad}:= \epsilon_T / (\epsilon_T + \epsilon_M)$.  It
should be noted that Eq.~\ref{eq:radeff} refers to the ``bolometric''
emission, i.e. it includes all frequencies for the whole duration of
the shell interaction. Since Earth-based observations have a limited
spectral and temporal coverage, the Eq.~\ref{eq:radeff} is only an
upper limit for radiative efficiencies inferred from actual
observations.
\begin{figure}
\includegraphics[width=8.5cm]{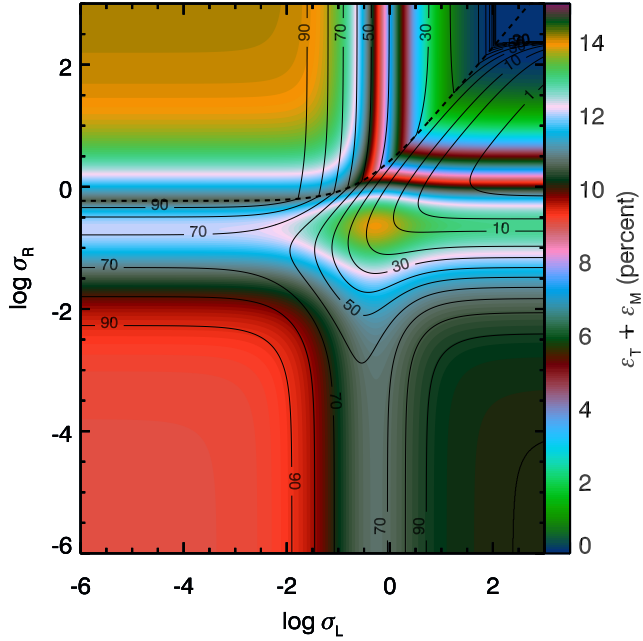}
\caption{Contours: $f_{\rm rad}$ (see Eq.~\ref{eq:radeff}) for
  different values of $(\sigma_L, \sigma_R)$. The contours indicate
  the following values of $f_{\rm rad}$ (per cent): $1, 5, 10, 20, 30,
  40, 50, 60, 70, 80, 90$, and $100$. In the region of the parameter
  space above the dashed line there is no FS, while the RS is always
  present for the considered parametrization (see also Fig.~1 of
  MA10). Filled contours: total dynamic efficiency $\epsilon_T +
  \epsilon_M$ in per cent. }
\label{fig:fr}
\end{figure}
Figure~\ref{fig:fr} shows that the radiative efficiency is not a
one-to-one map of the total dynamic efficiency. In particular, we note
that $f_{\rm rad}$ drops to under $10\%$ in the region $(\sigma_L>10,
\sigma_R>10)$. Furthermore, there is a region of maximal dynamic
efficiency for $\sigma_R \approx 0.2$ and $\sigma_L>1$, where the radiative
efficiency from internal shocks, which can only tap the thermal
energy in the regions downstream shocks, is not maximum. Nevertheless,
for small-to-moderate values of the magnetizations of both shells (lower
right quadrant of Fig.~\ref{fig:fr}), the radiative efficiency is a
good proxy of the total dynamical efficiency.

\section{Spectra and light curves of magnetized internal shocks in blazars}
\label{sec:emission}

Our aim is to produce synthetic spectra and light curves from our
numerical models of the interaction of two relativistic, magnetized
shells. With this purpose, we chose three models from our parameter
space, which are representative of different conditions that can be
encountered in blazar jets. The first model corresponds to a regime of
very low magnetization of both shells $(\sigma_L,\sigma_R)=(10^{-6},
10^{-6})$. The second and third models correspond to intermediate
$(10^{-2}, 10^{-2})$ and moderate/high shell
magnetizations $(1, 0.1)$.

\subsection{Average spectra}

\begin{figure}
\includegraphics[width=8.5cm]{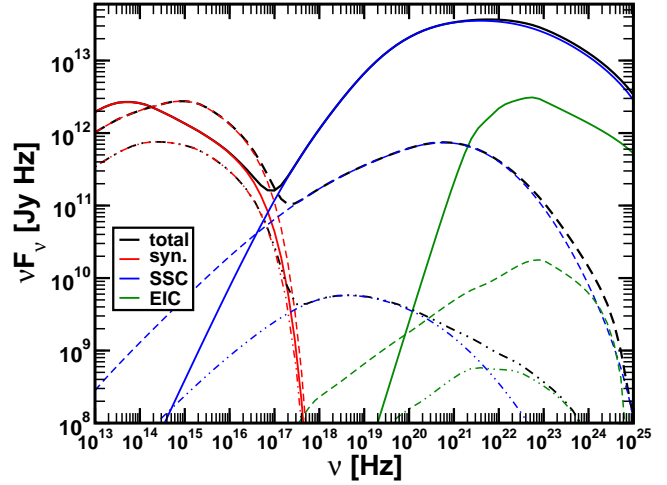}
\caption{Averaged spectra (time integration interval: $0$ - $100$ ks) for
  models $(10^{-6}, 10^{-6})$, $(10^{-2}, 10^{-2})$ and $(1, 0.1)$ (full,
  dashed, and dot-dot-dashed lines, respectively). Black colored lines
  show the total spectrum, while the red, blue and green lines show
  the contribution due to synchrotron, SSC and EIC emission,
  respectively.}
\label{fig:spec-comps}
\end{figure}

The spectrum of the weakly magnetized model $(10^{-6}, 10^{-6})$,
(Fig.~\ref{fig:spec-comps}; full lines) reproduces the typical
double-peaked spectrum of blazars. The synchrotron emission
(Fig.~\ref{fig:spec-comps}; solid red line) peaks at $4.6 \times
10^{13}\,$Hz, while the inverse-Compton (IC) emission, dominated by
the SSC component, peaks at $6.7 \times 10^{21}\,$Hz
(Fig.~\ref{fig:spec-comps}; solid blue line). In this case, the IC
spectral component is clearly dominating the overall spectrum.
At intermediate magnetizations (Fig.~\ref{fig:spec-comps}; dashed
lines) the synchrotron emission peaks at higher frequencies than in
the previous case, namely, at $7.5\times 10^{14}\,$Hz, while the IC
emission peaks at $6.0\times 10^{20}\,$Hz, as one would expect, since
a larger magnetic field increases the synchrotron peak frequency
\cite[e.g., ][]{Mimica:2004zz}. For these shell magnetizations, the
SSC also dominates the high energy emission, but now both the SSC and
EIC components are significantly weaker than in the model $(10^{-6},
10^{-6})$. Interestingly, moderate to high magnetizations
(Fig.~\ref{fig:spec-comps}; dot-dot-dashed lines) reduce the peak
frequency ($1.9\times 10^{14}\,$Hz) and substantially flatten the
synchrotron spectral component w.r.t. the intermediate magnetization
case. Furthermore, IC spectral components are notably weaker than the
synchrotron one for large magnetization, being the IC spectral peak
located at $4.2\times 10^{18}\,$Hz.

As can be seen from Fig.~\ref{fig:spec-comps}, the synchrotron
emission from all three models is of comparable intensity, while the
IC emission is much weaker in the strongly magnetized model $(1,
0.1)$. The reason for such a large difference in the high-energy
emission between the magnetized and the non-magnetized models lies in
the lower number density of emitting electrons (Eq.~\ref{eq:numdens})
and in the higher magnetic field of the magnetized model. The
magnetized model has a much lower number density due to its relatively
high $\sigma$, in both FS and RS emitting regions, which means that
there are less scatterers for the SSC and EIC processes. A high
magnetic field also means a reduction of the upper injection cut-off
(Eq.~\ref{eq:gamma2}), which in turn means that the seed synchrotron
photons in the SSC process are being up-scattered to lower frequencies,
explaining the small contribution of the SSC component to the
spectrum. The EIC component's upper cut-offs are determined by the
Klein-Nishina decline (see Sec.~\ref{sec:EIC}) and not by the upper
injection cut-off, which explains why the EIC spectral peaks of the
models are in a similar frequency range.
\begin{figure}
\includegraphics[width=8.5cm]{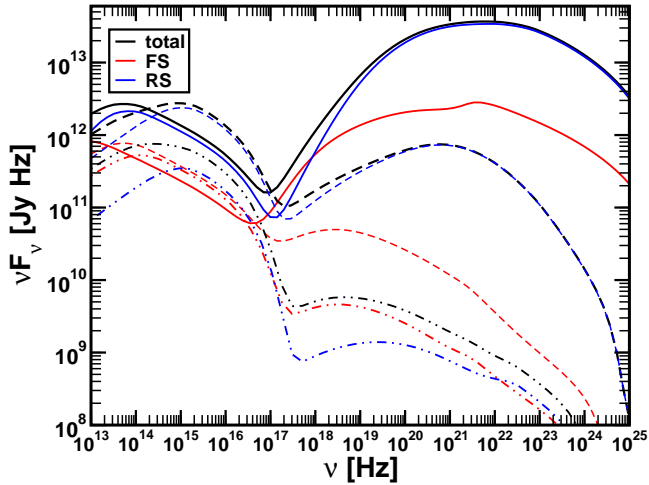}
\caption{Same as Fig.~\ref{fig:spec-comps}, but distinguishing the
  contributions of the FS (red lines) and of the RS (blue lines) to
  the total spectrum (black lines).}
\label{fig:spec-shocks}
\end{figure}

Figure~\ref{fig:spec-shocks} shows the contributions from the FS and
RS (red and blue lines, respectively) to the total spectrum (black
lines). Except close to the local minima between the two spectral
peaks, the RS contribution is dominant in the average spectra of the
models with low and intermediate magnetizations, $(10^{-6}, 10^{-6})$
and $(10^{-2}, 10^{-2})$ respectively. In the vicinity of the
aforementioned minima (located in the X-rays range), the FS
contribution tends to broaden the width of the minima and to soften
the spectral slope. At the moderate to high magnetizations of the
model $(1, 0.1)$ the FS is dominant except in the range $10^{15} -
10^{16}\,$Hz, where the FS and the RS have comparable contributions.
The reason is that the faster shell, through which the RS propagates,
is substantially more magnetized than the slower shell, so that the RS
has less particles to accelerate than the FS.

\subsection{Light curves}
\label{sec:lcs}

\begin{figure}
\includegraphics[width=8.5cm]{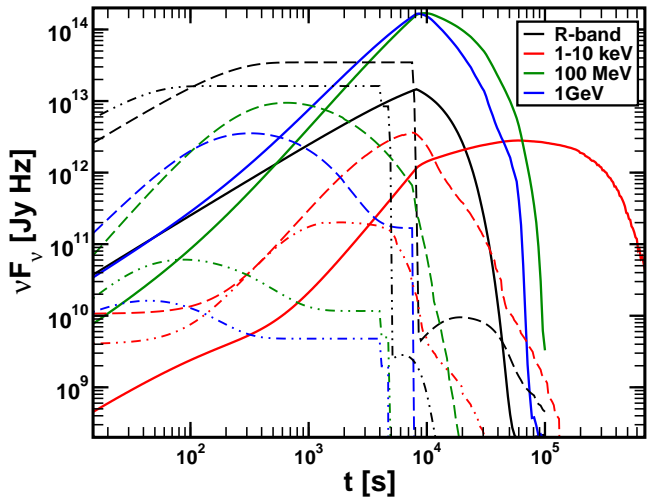}
\caption{Multi-wavelength light curves for the models $(10^{-6},
  10^{-6})$, $(10^{-2}, 10^{-2})$ and $(1, 0.1)$, shown with full,
  dashed and dot-dot-dashed lines, respectively. The R-band ($5\times
  10^{14}\,$Hz), X-ray band ($1-10\,$keV), as well as $0.1\,$GeV
  and $1\,$GeV light curves are shown in black, red, green and blue,
  respectively.}
\label{fig:mwlc}
\end{figure}

The multi-wavelength light curves of the models presented in the
previous section are displayed in Fig.~\ref{fig:mwlc}. We have picked
up several characteristic bands to analyse the data (R-band,
X-ray, $0.1\,$GeV and $1\,$GeV).

Comparing the R-band light curves of the three models we can see that
models $(10^{-2}, 10^{-2})$ and $(1, 0.1)$ exhibit properties of the
fast-cooling electrons emitting synchrotron radiation, while the model
$(10^{-6}, 10^{-6})$ shows the opposite, slow-cooling behavior. In
the latter case the maximum of the R-band light curve is reached when
the shocks cross the shells, and afterward the emission decays as the
particles cool down (no new particles are accelerated after both
shocks cross the shell). In the case of the model $(1, 0.1)$ one can
clearly notice two sudden drops in emission around 4\,ks, which
correspond to the moments when first the RS, and later, the FS cross
their respective shells. The almost vertical drops in emission are
indicative of a very efficient electron synchrotron-cooling
\footnote{\corr{Note that, since the high-latitude emission is
    ignored due to cylindrical geometry, the drops are too sharp and
    would be smoother were a conical jet geometry assumed.}}. At
intermediate magnetizations $(10^{-2}, 10^{-2})$ the first sharp drop
is observable as well, though here there is a weak late-time optical
emission between $10^4$ and $10^5$ seconds due to the SSC process.

\begin{figure}
\includegraphics[width=8.5cm]{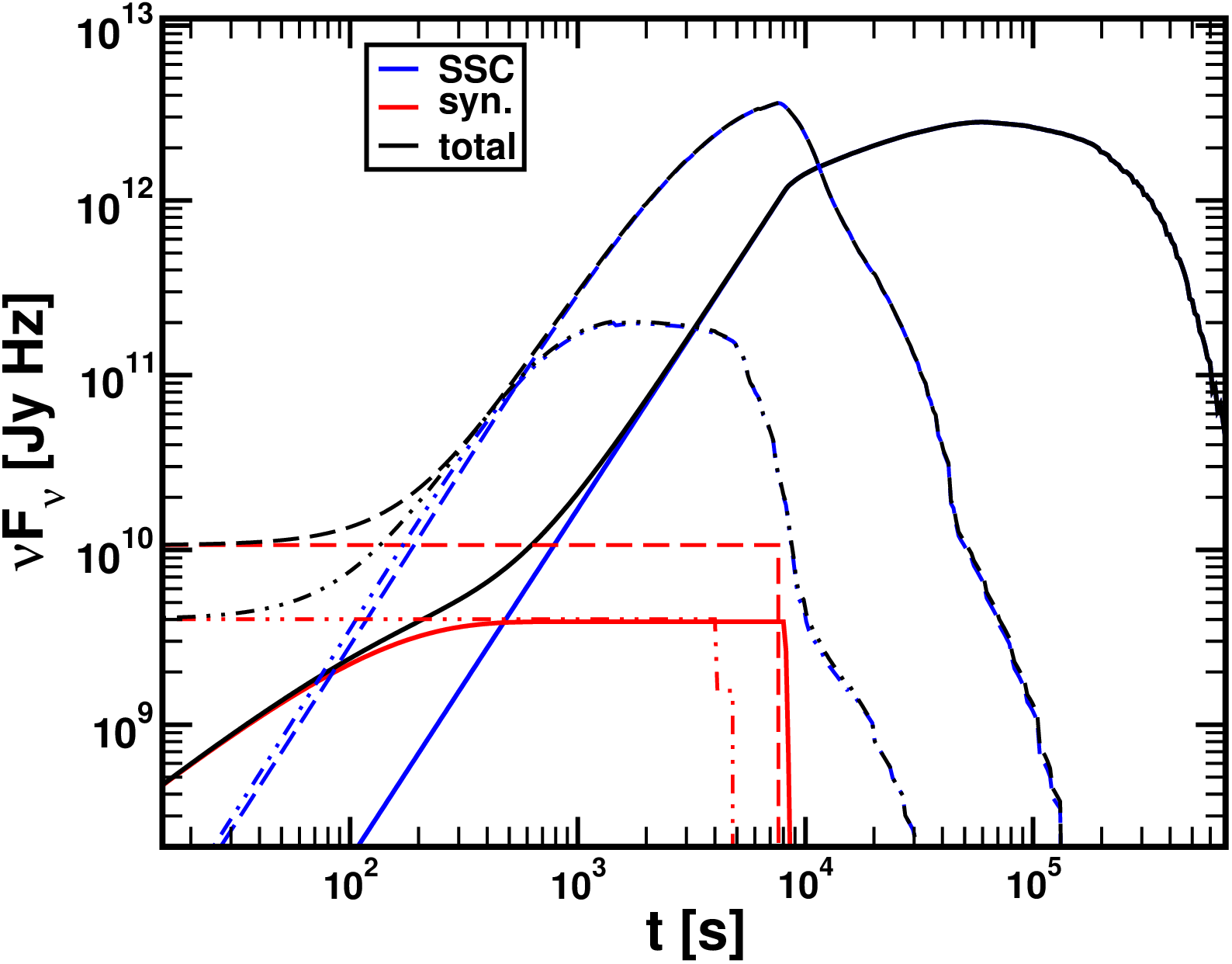}
\caption{X-ray light curves for the models $(10^{-6}, 10^{-6})$,
  $(10^{-2}, 10^{-2})$ and $(1, 0.1)$, shown with full, dashed and
  dot-dashed lines, respectively. The total light curve is shown in
  black, while the synchrotron and SSC contributions are shown in red
  and blue, respectively.}
\label{fig:X-ray}
\end{figure}
\begin{figure}
\includegraphics[width=8.5cm]{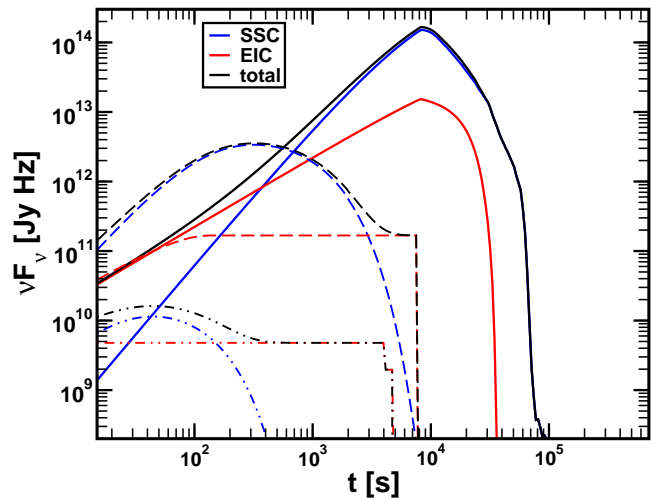}
\caption{$1\,$GeV light curves for the models $(10^{-6},
  10^{-6})$, $(10^{-2}, 10^{-2})$ and $(1, 0.1)$, shown with full,
  dashed and dot-dot-dashed lines, respectively. The total light curve is
  shown in black, while the EIC and SSC contributions are
  shown in red and blue, respectively.}
\label{fig:G}
\end{figure}
\begin{figure}
\includegraphics[width=8.5cm]{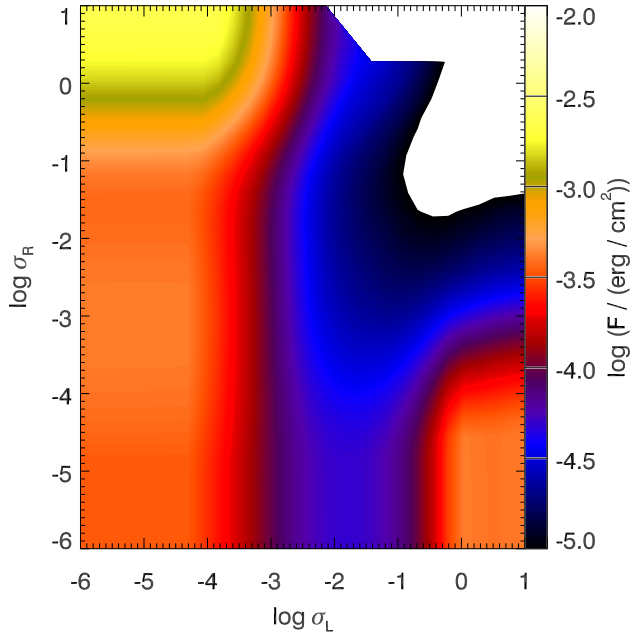}
\caption{Contours of the logarithm of the time ($0 - 100\,$ks) and
  frequency ($10^{12} - 10^{25}\,$Hz) integrated flux (i.e., fluence)
  as a function of the shell magnetization $\sigma_L$ and
  $\sigma_R$. Note that, different from Fig.~\ref{fig:fr}, the region
  of ultra-high magnetizations ($\sigma_{L,R}> 10$) is not included in
  this figure. The reason being that the computation of the integrated
  flux with such extreme magnetic fields requires a discretization of
  the two-dimensional integral in Eq.~\ref{eq:incint} in a very large
  number of intervals, making the calculation numerically
  impractical. Nevertheless, the trends at such high magnetizations
  can be easily extrapolated from the values displayed in the figure.}
\label{fig:prmlum}
\end{figure}
The emission in the X-ray band is a bit more involved, and to perform
a proper analysis we show in Fig.~\ref{fig:X-ray} both, the total
light curve (black lines), and the individual contributions to it of
the synchrotron and SSC processes (red and blue lines,
respectively)\footnote{We do not show the EIC light curve because its
  contribution is negligible at these frequencies.}. Except at very
early times, the emission is dominated by the SSC process in all
cases. The synchrotron emission in this band happens in an efficient
fast-cooling regime, which can be inferred from the fast drop of the
synchrotron components between 4 and 9\,ks. The fact that increasing
magnetic fields make that particles cool faster, explains that the
non-magnetized model peaks much later ($\simeq 60\,$ks) in this band
than the other two (more magnetized) models.

At energies of $1\,$GeV, there is only emission from IC processes
(Fig.~\ref{fig:G}). The model with the smaller magnetization displays
a clearly dominant EIC emission at early times, while in
the other two models EIC dominates the later times. In the models
$(10^{-2}, 10^{-2})$ and $(1, 0.1)$ EIC, similar to the synchrotron
emission in the X-ray band, sinks very quickly before 8\,ks,
indicating that the electrons are in a fast-cooling regime. In the
latter models, because of
the delays associated to the physical length of the emitting region,
the SSC contribution peaks very early and decays exponentially before
the sharp drop of the EIC emission (this is particularly the case of
the most magnetized model, in which the SSC component does not
significantly contribute to the light curve after $\simeq 400\,$s). In
contrast, the EIC emission of the model $(10^{-6}, 10^{-6})$ shows a
much more prominent peak and a shallower decay from the maximum (at
$\simeq 9\,$ks), both features being characteristics from electrons in
a slow-cooling regime.

\section{Global parameter study}
\label{sec:params}

\begin{figure*}
\includegraphics[width=8.5cm]{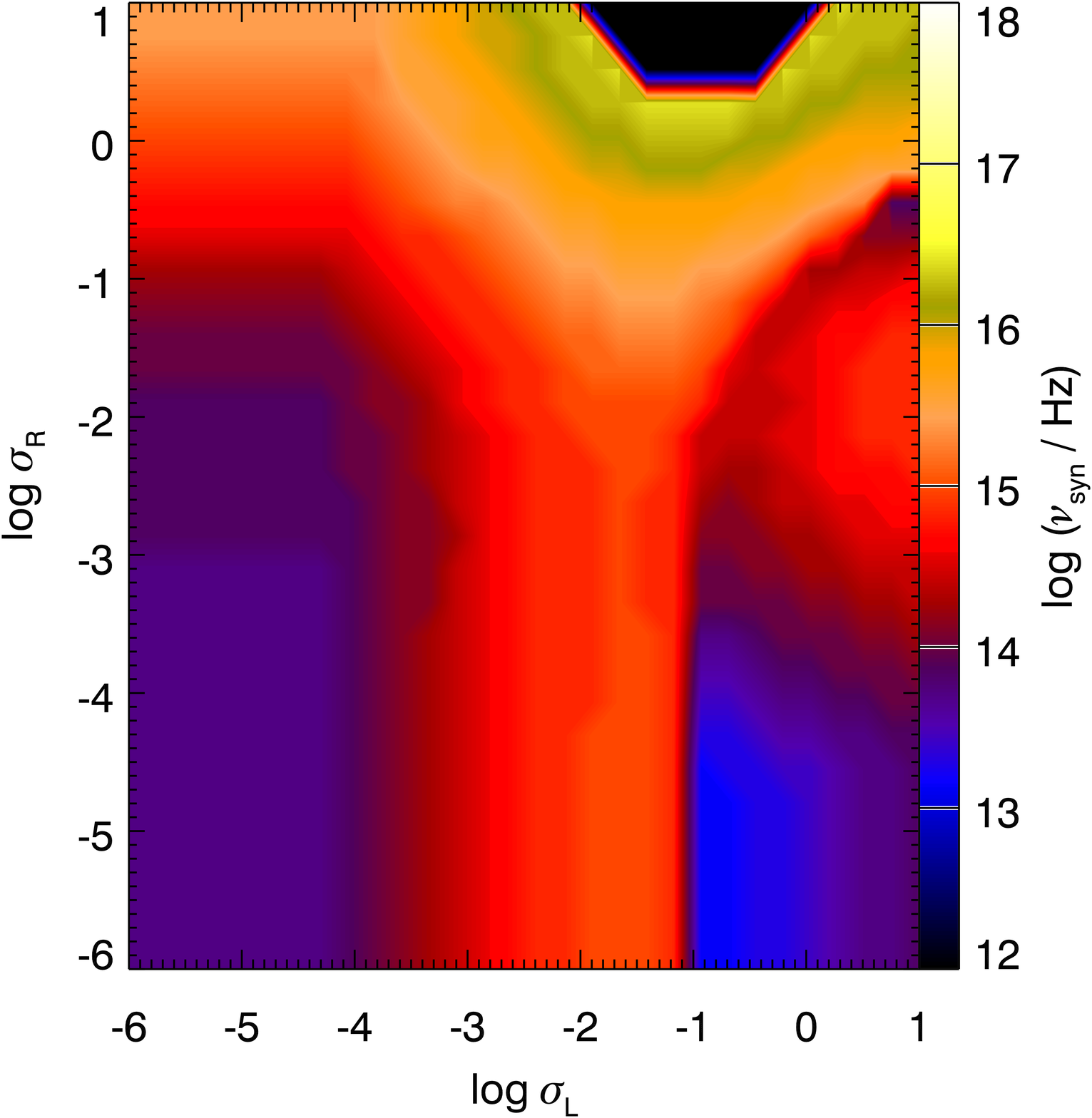}\hspace{0.2cm}
\includegraphics[width=8.5cm]{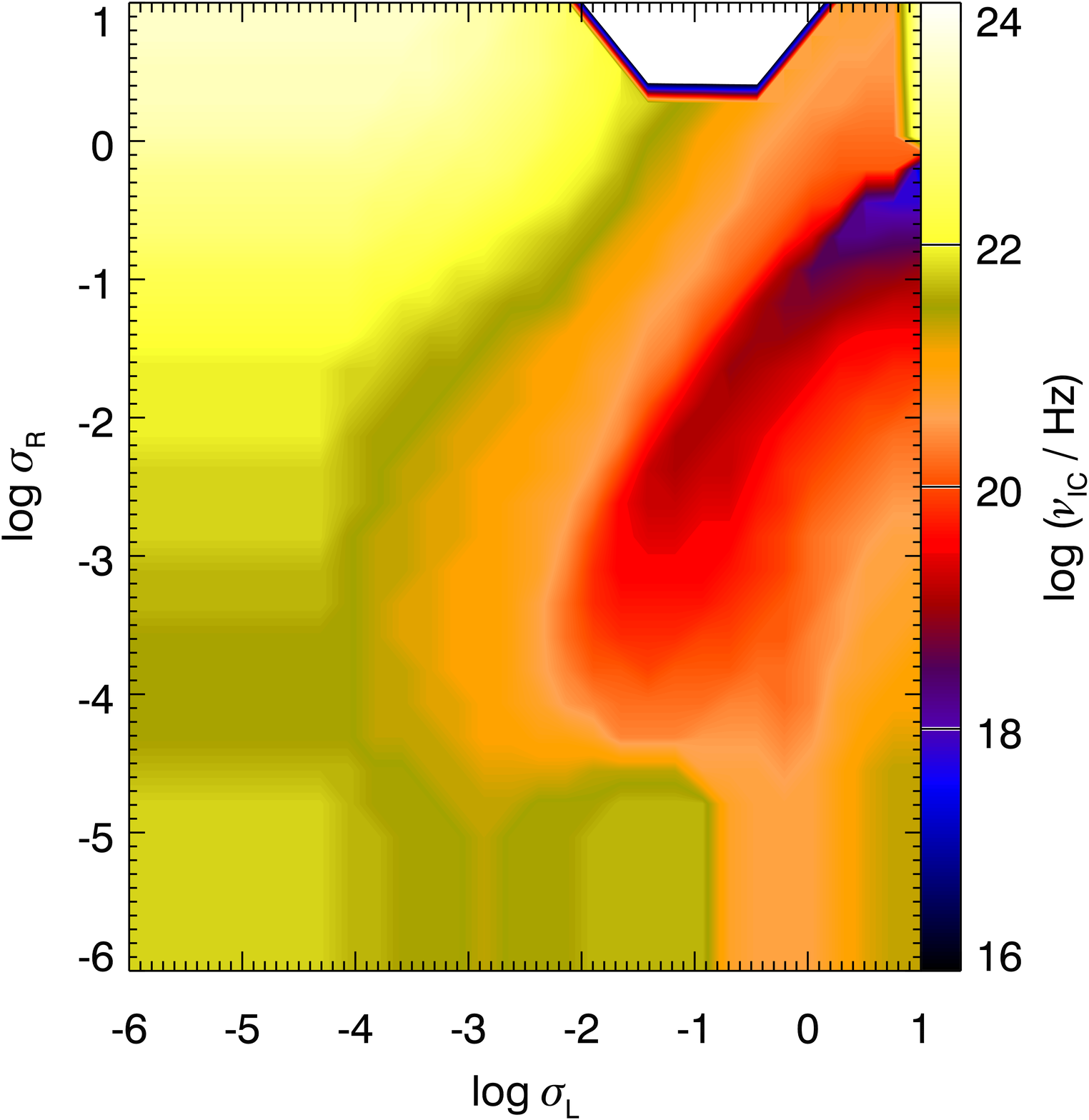}
\caption{Contours of the frequency of the spectral maxima of the
  synchrotron (left panel) and of the inverse Compton (right panel)
  emission as a function of $\sigma_L$ and $\sigma_R$.}
\label{fig:prmfreq}
\end{figure*}
\begin{figure*}
\includegraphics[width=8.5cm]{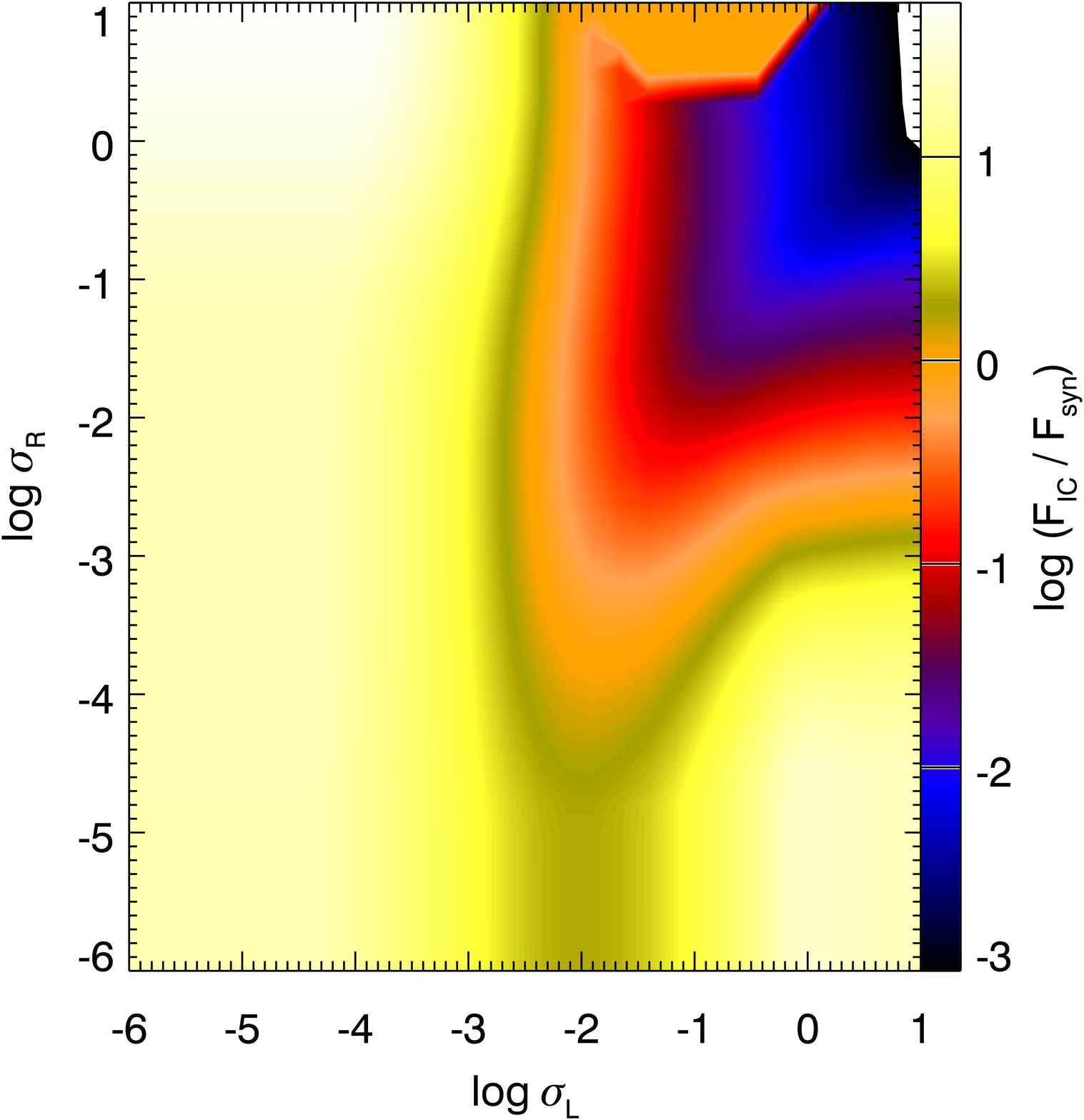}\hspace{0.2cm}
\includegraphics[width=8.5cm]{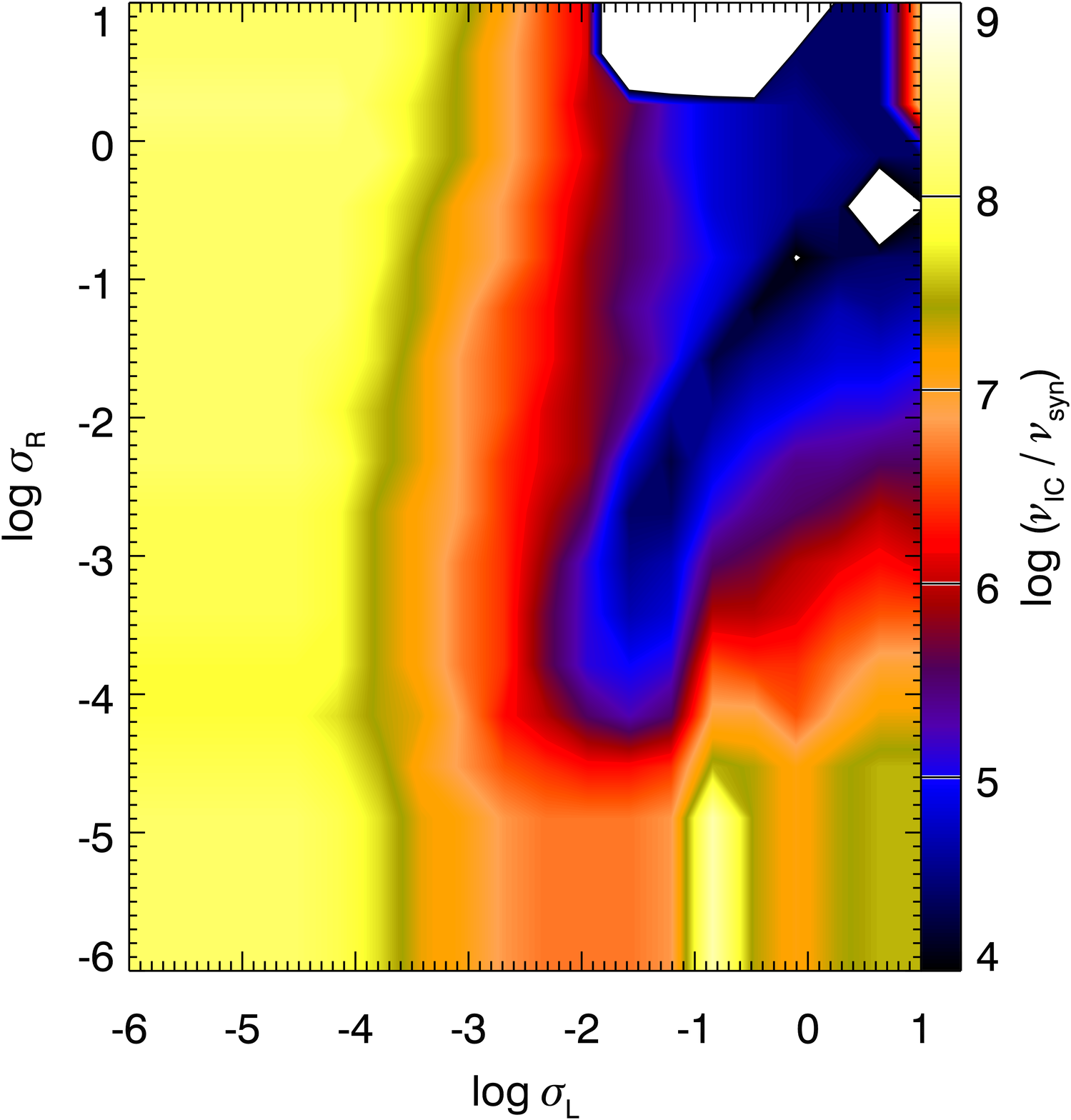}
\caption{Left panel: contours of the logarithm of the ratio of the IC
  and synchrotron fluence as a function of the shell magnetization
  $\sigma_L$ and $\sigma_R$. Right panel: same as left panel, but for
  the ratio of the frequency of the spectral maxima of the synchrotron
  and inverse Compton emission.}
\label{fig:fluxrat}
\end{figure*}
In the following we present the results of the global parameter study
of the dependence of the emitted radiation on the shell
magnetization\footnote{All two-dimensional plots in this section have
  been produced using a logarithmically spaced grid of $30\times 30$
  in the $\sigma_L\times\sigma_R$ parameter space. For each of the
  points we computed light curves on $96$ logarithmically spaced
  frequencies (between $10^{12}$ Hz and $3\times 10^{25}$ Hz) for
  $120$ logarithmically spaced points in time (between $2$ and $10^5$
  seconds). A finer coverage of the $\sigma_L\times\sigma_R$ parameter
  space was not practical due to the prohibitively high memory and
  computational time requirements on the available
  machines.}. Figure~\ref{fig:prmlum} shows the fluence as a function
of $\sigma_L$ and $\sigma_R$.  We can see that, as expected, the
fluence roughly follows $f_R$ (see Sect.~\ref{sec:efficiency}). The
region with most luminous internal shocks (upper left corner of
Fig.~\ref{fig:prmlum}) happens for a moderately-to-strongly magnetized
slow shell and a weakly magnetized fast shell, whereby the FS does not
exist. The emission weakens as the magnetization of the fast shell
increases, with the exception of the region where the fast shell is
strongly magnetized but the slow shell is weakly magnetized (lower
right corner of Fig.~\ref{fig:prmlum}). We conclude that, as was
indicated in the Section 4.4 of MA10, a large difference in the
magnetization of the shells yields stronger dissipation and more
luminous internal shock(s) than when both shells are weakly
magnetized.

Figure~\ref{fig:prmfreq} shows the spectral maxima of the synchrotron,
$\nu_{\rm max, syn}$, and the inverse Compton emission, $\nu_{\rm max,
  IC}$ (left and right panels, respectively). From
Fig.~\ref{fig:spec-comps}, one could anticipate a trend we confirm
here, with the parametric space coverage, namely, that the IC emission
is more sensitive to changes in the magnetization than the
synchrotron emission. This statement reflects itself in
Fig.~\ref{fig:prmfreq} through the fact that the range of variation of
$\nu_{\rm max, IC}$ is larger than that of $\nu_{\rm max, syn}$. Thus,
the IC spectral peak becomes a better proxy of the magnetization of
the shells than the synchrotron peak. Except at small shell
magnetizations, the IC emission happens in a fast-cooling regime, and
the dependence of $\nu_{\rm max, IC}$ with the magnetization is
similar to that of $f_{\rm rad}$. Complementary, at small shell
magnetizations, the map of $\nu_{\rm max, syn}$, resembles very much
to that of $f_{\rm rad}$ (compare the lower half of Figs.~\ref{fig:fr}
and \ref{fig:prmfreq}-left).

The left panel of the figure~\ref{fig:fluxrat} shows the ratio of the
IC to synchrotron fluence. The trend is quite similar to that of the
integrated flux shown in Fig.~\ref{fig:prmlum}. When both shells are
strongly magnetized ($\sigma\simmore 0.1$) the IC emission drops
significantly. In the region $\sigma_L\simless 10^{-3}$ the ratio is
between a unity and $\simeq 60$, with a similar behavior in the region
$(\sigma_L\simmore 0.1, \sigma_R\simless 0.1)$. The region of low
radiative efficiency around $\sigma_L\simeq 0.01$ appears as a dark
vertical band in the plot ($\sigma_R\simless 10^{-4}$, where both
synchrotron and IC processes provide a similar fluence. The right
panel of Fig.~\ref{fig:fluxrat} shows the ratio of the frequencies of
the IC and synchrotron spectral maxima shown in
Fig.~\ref{fig:prmfreq}. From an observational point of view, the
fluence might be much more robust and significant than the peak IC
and synchrotron frequencies, which can be difficult to
measure. However, for the ratio $\nu_{\rm IC}/\nu_{\rm syn}$, the
lower right and upper left corners of the plot display noticeably
different values. Thus, one can use the frequency ratio together with
the fluence ratio in order to break the degeneration in the fluence
ratio when one of the shell is very magnetized and the other is not
magnetized (see Sec.~\ref{sec:dis_glob} for further discussion of this
point).

\section{Discussion and summary}
\label{sec:discussion}

We have extended the study of the dynamic efficiency performed in MA10
by computing the multi-wavelength, time-dependent emission from
internal shocks. In this section we discuss and summarize our findings.

\subsection{Emission mechanisms and magnetization}
\label{sec:dis_emiss}
In Section~\ref{sec:emission} we show the average spectra and
multi-wavelength light curves of three typical models from the
parameter space considered in this paper. Synchrotron emission
dominates for $\nu \simless 10^{17}$ Hz, and is rather independent of
the shell magnetization (Fig.~\ref{fig:spec-comps}). The RS dominates
synchrotron emission for weakly magnetized shells, while in the case
of strongly magnetized shells the FS and RS have comparable
contributions. R-band light curves (black lines on
Fig.~\ref{fig:mwlc}) show that the synchrotron emission is due to the
slow-cooling electrons only for the weakly magnetized model, while for
shells with $\sigma \simmore 0.01$ electrons are fast-cooling.

The SSC emission dominates in the X-ray band and higher frequencies
(Fig.~\ref{fig:spec-comps}). However, at early times the synchrotron
emission dominates in X-rays (Fig.~\ref{fig:X-ray}), while in
$\gamma$-rays the situation is more complex. For the weakly magnetized
model (slow-cooling electrons), EIC dominates the early emission,
while in the moderate-to-highly magnetized models EIC dominates the
late-time emission. The reason for this is that, in the magnetized
models, the high-energy tail of the electron distribution disappears
very quickly, so that the incoming synchrotron photons cannot be
up-scattered into the $1$\,GeV range. In the weakly magnetized models
there are enough slow-cooling electrons at sufficiently high energies
for the SSC to dominate over EIC at later times.

Finally, from Fig.~\ref{fig:spec-comps} we see that the IC emission is
weaker the more magnetized the shells. This is due to the requirement
of our model that the shell luminosity (Eq.~\ref{eq:luminosity}) be
constant regardless of $\sigma$. From Eq.~\ref{eq:numdens} we see that
for $\sigma\gg 1$ the number density in the shells behaves as $\approx
\sigma^{-1}$. Since the IC emission depends on the number of electrons
(EIC linearly and SSC quadratically), it is clear that the IC emission
muss necessarily drop for large $\sigma$.  From the analysis of the
three representative models, we conclude that the shell magnetization
imprints two main features on the emission properties of blazars. On
the one hand, the magnetization changes the ratio of integrated flux
below the synchrotron peak to the integrated flux below the
IC-dominated part of the spectrum. On the other hand, the
magnetization determines whether electrons are slow-cooling (for
weakly magnetized shells) or fast-cooling (moderate-to-high
magnetization).

\subsection{Global trends}
\label{sec:dis_glob}

We performed a global parameter study (Section~\ref{sec:params}) to
investigate the dependence of some observational quantities on the
shell magnetization. As discussed in Section~\ref{sec:params}, the
integrated flux (Fig.~\ref{fig:prmlum}) follows the trend already
shown by the radiative efficiency (Fig.~\ref{fig:fr}). However, the
integrated flux and the spectral maxima are quantities dependent on
the particular values we have taken in our model, specifically, on the
physical size of the shells and their bulk Lorentz factors, as well as
on the source redshift. On the other hand, in MA10 we show that the
dynamic efficiency is very weakly dependent on the shell bulk Lorentz
factor, i.e. it only depends on the shell magnetization for a fixed
$\Delta g$. In order to eliminate the dependence on absolute
quantities in Fig.~\ref{fig:fluxrat} we show IC-to-synchrotron flux
ratio, as well as the ratio of the frequency of IC to the synchrotron
spectral maxima. The shape of the contours on the left panel of
Fig.~\ref{fig:fluxrat} does not exactly follow the one in
Fig.~\ref{fig:fr}: there is a much stronger dependence on $\sigma_L$
than on $\sigma_R$ in most of the scanned parameter
space. Nonetheless, in the lower half of the plots, the behavior of
both $F_{\rm IC}/F_{\rm syn}$ and $f_{\rm R}$ is similar. For example,
if we keep $\sigma_R$ constant and equal to $10^{-6}$ and vary
$\sigma_R$, we note that the radiative efficiency is larger than 90\%
for $\sigma_L<0.01$, then it decays to a local minimum, and
successively grows again to reach values in excess of 90\% ($\sigma_L
\simmore 1$). Comparatively, at small values of $\sigma_R$ the ratio
$F_{\rm IC}/F_{\rm syn}$ is close to its maximum for $0.01\simless
\sigma_L\simmore 1$, and touches a minimum in the same interval as
$f_{\rm R}$. The upper half of Figs.~\ref{fig:fluxrat} (left) and
\ref{fig:fr} does not show the same qualitative behavior. The reason
for this discrepancy is that RS dominates the emission, and thus the
overall radiative properties are more sensitive to the magnetization
of the fast shell through which it propagates.

Interestingly, there is a certain degree of degeneration in the values
both of the radiative efficiency and of the $F_{\rm IC}/F_{\rm syn}$
ratio considering the regions where one of the two shells is very
magnetized and the other is basically non-magnetized (i.e., the upper
left and lower right corners of Figs.~\ref{fig:fluxrat} and
\ref{fig:fr}). In both cases, the radiative efficiency and the fluence
ratio are close to their respective maximum values. However, we can
distinguish between the case of high-$\sigma_L$/low-$\sigma_R$ and the
case of low-$\sigma_L$/high-$\sigma_R$ by looking at the ratio of peak
frequencies $\nu_{\rm IC}/\nu_{\rm syn}$ (right panel of
Fig.~\ref{fig:fluxrat}). A noticeably smaller $\nu_{\rm IC}/\nu_{\rm
  syn}$ ratio corresponds to the former case than to the later.

\begin{figure*}
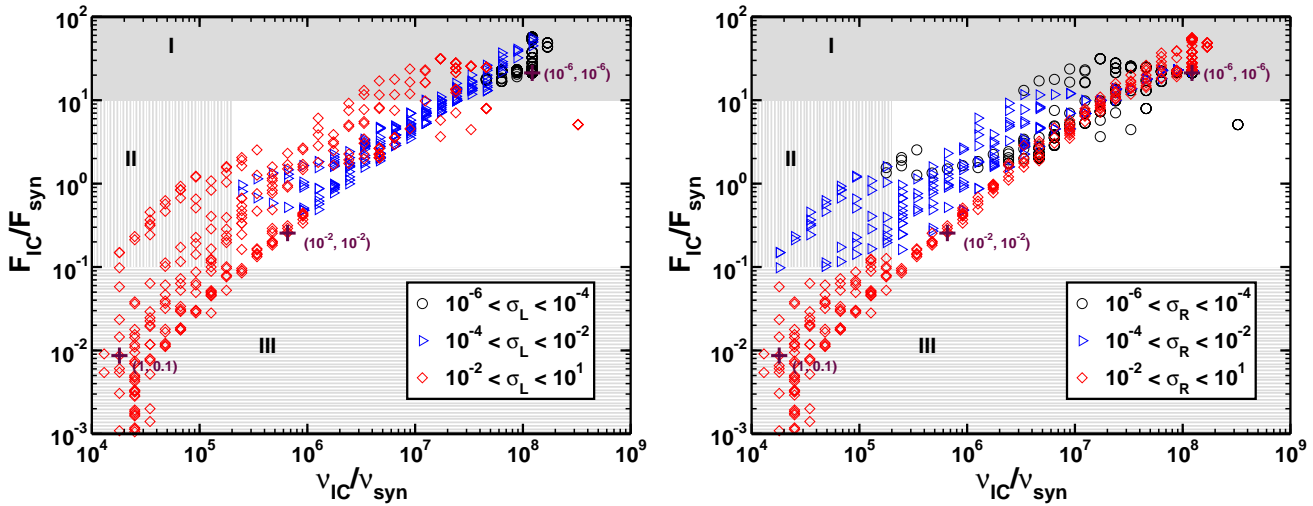

\includegraphics[width=8.5cm]{ratios_L.eps}\hspace{0.2cm}
\includegraphics[width=8.5cm]{ratios_R.eps}
\caption{Left panel: $F_{\rm IC}/F_{\rm syn}$ (ratio of the IC to the
  synchrotron fluence) as a function of $\nu_{\rm IC}/\nu_{\rm syn}$
  (ratio of the maximum spectral frequencies in IC and synchrotron
  ranges) for the models considered in Sec.~\ref{sec:params}. The
  models have been grouped in three bins according to $\sigma_L$ and
  are annotated with black circles ($10^{-6}<\sigma_L<10^{-4}$), blue
  triangles ($10^{-4}<\sigma_L<10^{-2}$) and red diamonds
  ($10^{-2}<\sigma_L<10^1$). Right panel: same as left panel, but in
  this case, the models have been grouped according to $\sigma_R$ in
  the same bins and same coloring and symbols in as the left
  panel. Shaded areas denote three regions of interest (see text for
  details). The three reference models in this paper are marked with
  purple crosses, and their respective magnetizations are overlaid.}
\label{fig:ratios}
\end{figure*}

The previous analysis suggests that with the combined information of
the fluence and peak-frequency ratios, one could try to figure out, by
using observational data, which is the rough magnetization of the
shells of plasma whose interaction yields flares in blazars. To serve
such a purpose, we display in Fig.~\ref{fig:ratios} our models in a 2D
parameter space whose horizontal and vertical directions are $\nu_{\rm
  IC}/\nu_{\rm syn}$ and $F_{\rm IC}/F_{\rm syn}$, respectively. We
notice that the computed models are distributed in a broad region
which, nevertheless, shows a relatively tight correlation between
$F_{\rm IC}/F_{\rm syn}$ and $\nu_{\rm IC}/\nu_{\rm syn}$. In the left
panel of Fig.~\ref{fig:ratios}, we display our models in three
different colors according to the magnetization of the left shell. The
same has been done in the right panel, but for the right shell. 

Based on the degree of variation of magnetization between the fast
and the slow shells, we have divided the parameter space in three broad regions
(labeled with roman numerals {\bf I}, {\bf II} and {\bf III} in
Fig.\ref{fig:ratios}), where the shells have the following
characteristics:
\begin{enumerate}[\bf I:]
\item moderately magnetized fast shell colliding with a
   weakly magnetized slow shell, or weakly magnetized fast shell
   interacting with a strongly magnetized slow one;
\item strongly magnetized fast and moderately magnetized slow shells;
\item strongly magnetized fast and slow shells.
\end{enumerate}
The first thing to note is that for models in the region {\bf III}
both the IC emission and its frequency maximum are lower compared to
the rest of the models. This leads us to the conclusion that when the
flow is strongly magnetized \emph{and} the magnetization does not vary
substantially the IC signature is expected to be relatively
weak. Furthermore, region {\bf II} shows that in the case of a larger
variation in magnetization (i.e., weakly magnetized slow shell) the
frequency maximum remains low, but the IC signature becomes
substantially higher. Finally, in the region {\bf I} we see that when
the variation in magnetization is more extreme (i.e. a collision of a
weakly and a strongly magnetized shells) we get a very strong IC
signature and its frequency maximum is shifted to much higher
energies.

\subsection{Radiative and dynamic efficiency}
\label{sec:dis_eff}

As discussed in Section~\ref{sec:efficiency}, the radiative efficiency
$\epsilon_e f_{\rm rad} (\epsilon_T + \epsilon_M)$ does not have a
one-to-one correspondence to the dynamic efficiency $(\epsilon_T +
\epsilon_M)$. While the latter peaks in the region $\sigma_L \approx
1$ and $\sigma_R \approx 0.2$, the former reaches its maximum in the
region $(\sigma_L\simless 10^{-4}, \sigma_R \simmore 10)$. The same
can be concluded from the time- and frequency-integrated flux shown in
Fig.~\ref{fig:prmlum}. For purposes of the rest of this discussion we
will use $f_{\rm rad}$ as a proxy for the radiative efficiency. 

We note that in the region of maximum $f_{\rm rad}$ the FS does not
exist. However, we see another region of high $f_{\rm rad}$ in the
opposite corner of Fig.~\ref{fig:fr} the efficiency is quite high as
well. Consistent with the discussion in the previous subsection and
with what is shown in region {\bf I} in Fig.~\ref{fig:ratios}, we
conclude that the radiative efficiency is maximal when the variation
in magnetization between the colliding shells is large.

\subsection{Conclusions and future work}
\label{sec:dis_con}

Under the assumption of a constant flow luminosity we find that there
is a clear difference between the models where both shells are weakly
magnetized ($\sigma\simless 10^{-2}$) and those where, at least, one
shell has a $\sigma\simmore 10^{-2}$. We obtain that the radiative
efficiency is largest in those models where, regardless of the
ordering, there is a large variation in the magnetization of the
interacting shells. Furthermore, substantial differences between
weakly and strongly magnetized shell collisions are observed in the
inverse-Compton part of the spectrum, as well as in the optical, X-ray
and $1$~GeV light curves.

In the previous sections we have deepened our analysis of the
radiative efficiency of the process of collision of magnetized
relativistic shells of plasma. We have studied this problem from a
mostly theoretical point of view, where the intrinsic properties of
the flow (in particular the magnetization) have been related to the
properties of the resulting (synthetic) spectra and light curves. It
is, however, worthwhile to provide suitable links between our
theoretical results and the observed properties of blazar
flares. Thus, we propose a way to distinguish observationally between
weakly magnetized from magnetized internal shocks by comparing the
maximum frequency of the inverse-Compton and synchrotron part of the
spectrum to the ratio of the inverse-Compton and synchrotron fluence.

For a given flare taken in isolation, our model may predict which is
the range of magnetizations, which have to be invoked to fit the
spectrum. However, such a model fitting is not fully satisfactory,
since it is strongly dependent on the details of the theoretical
model. A more generic knowledge of the physical conditions in the
flaring regions can be obtained by arranging the observational data in
plots where the fluence ratio $F_{\rm IC}/F_{\rm syn}$ is represented
versus $\nu_{\rm IC}/\nu_{\rm syn}$. The reason being that the fluence
and frequency ratios are redshift and \emph{source} independent, since
they are mostly influenced by the variation of the bulk magnetization
of the blazar jets (assuming that the viewing angle is fixed).  We
note that different flares of the same blazar, as well as different
flares of distinct blazars can be plotted in such graphs and compared
with our theoretical predictions. In addition, an average over a
number of flares of the same source might also be interpreted using
our model if the magnetization ratio of different pairs of colliding
shells is similar.

Our results suggest that the variability in the flow magnetization is
a factor that shall be considered to explain the observed continuity
of properties of the blazar sequence
\citep[e.g.,][]{Fossati:1998ay,Ghisellini:1998hg,Ghisellini:2008ax}. Looking
at Figs.~\ref{fig:spec-comps} and~\ref{fig:spec-shocks}, it is evident
that if the magnetization of the shells is not too large
($\sigma_{L,R}\simless 10^{-2}$), increasing the flow magnetization
shifts the synchrotron peak towards the UV band and lowers the IC
peak. Leaving aside orientation effects (which we are neglecting here
since we fix the viewing angle), such a behavior suggests that LBL
blazars may correspond to barely magnetized flows, while HBL blazars
could correspond to moderately magnetized ones. If the magnetization
is large ($\sigma_{L,R}\simmore 0.1$) the synchrotron peak shifts
towards lower frequencies, and the IC spectral peak falls three orders
of magnitude below the synchrotron peak. The latter situation seems
not to be observed and, thus, we conclude that this is an indication
that the typical value of the magnetization in the flow of blazars is
$\sigma\simless 10^{-2}$. We note that this value is about one order
of magnitude smaller than that suggested for the flow in gamma-ray
bursts \citep[e.g.,][]{Giannios:2006xt}.

\corr{Our results are not in contradiction with the common view,
  according to which, the variation of the properties of the blazar
  family correspond to the changes in the bolometric luminosity of the
  synchrotron component ($L_{\rm bol, sync}$;
  \citealt{Fossati:1997ay}). In such a case LBLs and HBLs are extrema
  of a one-parameter family, where LBLs are more radio luminous than
  HBLs. However, $L_{\rm bol, sync}$ is not a "direct" physical
  property of the plasma in a relativistic jet, since the same $L_{\rm
    bol, sync}$ may arise with an infinite number of combinations of
  bulk Lorentz factors, blazar orientations with respect to the line
  of sight, flow magnetization, etc. In this paper, we explore the
  role that the flow magnetization (a direct physical property of the
  emitting plasma) plays in shaping not only $L_{\rm bol, sync}$, but
  also the whole spectrum. We conclude that the flow magnetization
  alone is enough to explain the difference in $L_{\rm bol, sync}$ and
  in the high-energy part of the spectrum (i.e., in the
  Compton-dominated regime) found in the blazar family. We also point
  out that \cite{Fossati:1998ay} also arrive to the same qualitative
  conclusion, since they explain that fixing the bulk Lorentz factor,
  and assuming that the SSC model be valid for all sources, the
  spectral differences in the blazar sequence shall result from a
  systematic variation in magnetic field strength, HBLs having the
  highest random field intensity. We go in this paper an step
  further. When considering the dynamical changes induced by
  non-negligible {\em macroscopic} magnetic fields, the same
  conclusion as in \cite{Fossati:1998ay} holds, but we remark that, in
  our case, the total jet luminosity is kept constant (by construction
  of our models), only the magnetization is varied.}

 \corr{A problem with the internal shock scenario is that
  it is apparently not able to explain the ultra-fast variability of
  TeV blazars \citep[e.g.,][]{Albert:2007aa,Aharonian:2007aa}. In order
  to properly account for this fast variability (on the time scales of
  minutes) fast ``minijets'' \citep{Giannios:2009zi} or
  ``spines/needles'' \citep{Tavecchio:2008aa} need to exist in a much
  larger, slower jet. We do not consider minijets in our current
  models, although they will need to be considered in the future.}

In the future work we will improve our modeling by including the
resistive dissipation as a source of energy for the radiating
non-thermal particles. This will make it possible to asess whether
radiative and dynamical efficiencies have a one-to-one correspondence
or if there is a fundamental degree of independence. In the latter
case it might be difficult to infer the flow properties except in
those asymptotic cases where the dissipation either does not play a
significant role or it dominates the dynamics. Furthermore, we will
study how changing the viewing angle and the Doppler factor reflect on
the observational signature. \corr{Finally, we will include the
  effects of the presence of a thermal component during particle
  injection at very magnetized shocks
  \citep{Sironi:2009aa,Giannios:2009zm}}.

\section*{Acknowledgments}
\corr{We thank the anonymous referee for the constructive criticism
  and suggestions for the improvement of this paper.} 
The authors are grateful to Eduardo Ros \corr{and to M. Joshi} for valuable comments and
discussion. We also acknowledge the support from the European
Research Council (grant CAMAP-259276), and the partial support of
grants AYA2010-21097-C03-01, CSD2007-00050, and PROMETEO-2009-103. The
authors thankfully acknowledge the computer resources, technical
expertise and assistance provided by the Barcelona Supercomputing
Center - Centro Nacional de Supercomputaci\'on.

\appendix

\section{Comparison of the dynamic efficiency definitions}
\label{app:dyneff}

In this appendix we compare two alternative definitions of the dynamic
efficiency, one due to MA10 and the other inspired by
\citet[][NKT11 in the following]{Narayan:2011df}.

MA10 define in their Eq.~12 three components (kinetic, thermal and
magnetic, respectively) of the total energy in each shell:
\begin{equation}\label{eq:energiesMA10}
  \begin{array}{rcl}
    E_K(\Gamma, \rho, \Delta x) & := & \Gamma (\Gamma - 1) \rho c^2 \Delta x\\[4mm]
    E_T (\Gamma, \rho, p, \Delta x)& := & [(\rho \varepsilon + p) \Gamma^2 - p] \Delta x\\[4mm]
    E_M (\Gamma, \rho, \sigma, \Delta x) & := & \left(\Gamma^2 -
      \dsfrac{1}{2}\right)\rho \sigma c^2 \Delta x
  \end{array}\, ,
\end{equation}
where $\rho$, $p$, $\varepsilon$, $\sigma$ and $\Gamma$ are the fluid
rest-mass density, thermal pressure, specific internal energy density,
magnetization parameter and the Lorentz factor. $\Delta x$ is the
width of the shell. Then, the total dynamical efficiency is defined as
the ratio of the sum of the magnetic and thermal energies in the
post-shock state to the total initial energy (MA10). 

Recently, NKT11 have computed the radiative efficiency of magnetized
internal shocks using a slightly different definition. The definition
of NKT11 has the advantage of resulting into a radiative efficiency,
which is Lorentz-invariant, while the one of MA10 is not. Therefore,
the dynamic efficiency can be computed in the lab frame and then used
in the rest frame of the contact discontinuity to be directly compared
with the radiative efficiency, as was done in
Sec.~\ref{sec:efficiency}. However, NKT11 approach does not shed any
light on the problem of obtaining a true energetic efficiency in terms
of the initial conditions of the shells. This is because their
definition does not consider the efficiency of conversion of the {\em
  initial} total energy of the shells to radiation but, instead, the
efficiency of conversion of the enthalpy per particle {\em after} the
shell collision into radiation enthalpy.

With the aim of introducing a Lorentz invariant, energy-based
definition of the radiative efficiency in terms of a quantity akin to
the dynamic efficiency of MA10, we consider the following expressions
for the kinetic, thermal and magnetic energies

\begin{equation}\label{eq:energiesNKT11}
  \begin{array}{rcl}
    \hat{E}_K(\Gamma, \rho, \Delta x) & := & \Gamma^2 \rho c^2 \Delta x\\[4mm]
    \hat{E}_T (\Gamma, \rho, p, \Delta x)& := & \Gamma^2 (\rho \varepsilon + p) \Delta x\\[4mm]
    \hat{E}_M (\Gamma, \rho, \sigma, \Delta x) & := & \Gamma^2\rho
    \sigma c^2 \Delta x
  \end{array}\, .
\end{equation}
%

\begin{figure*}
\includegraphics[width=8.5cm]{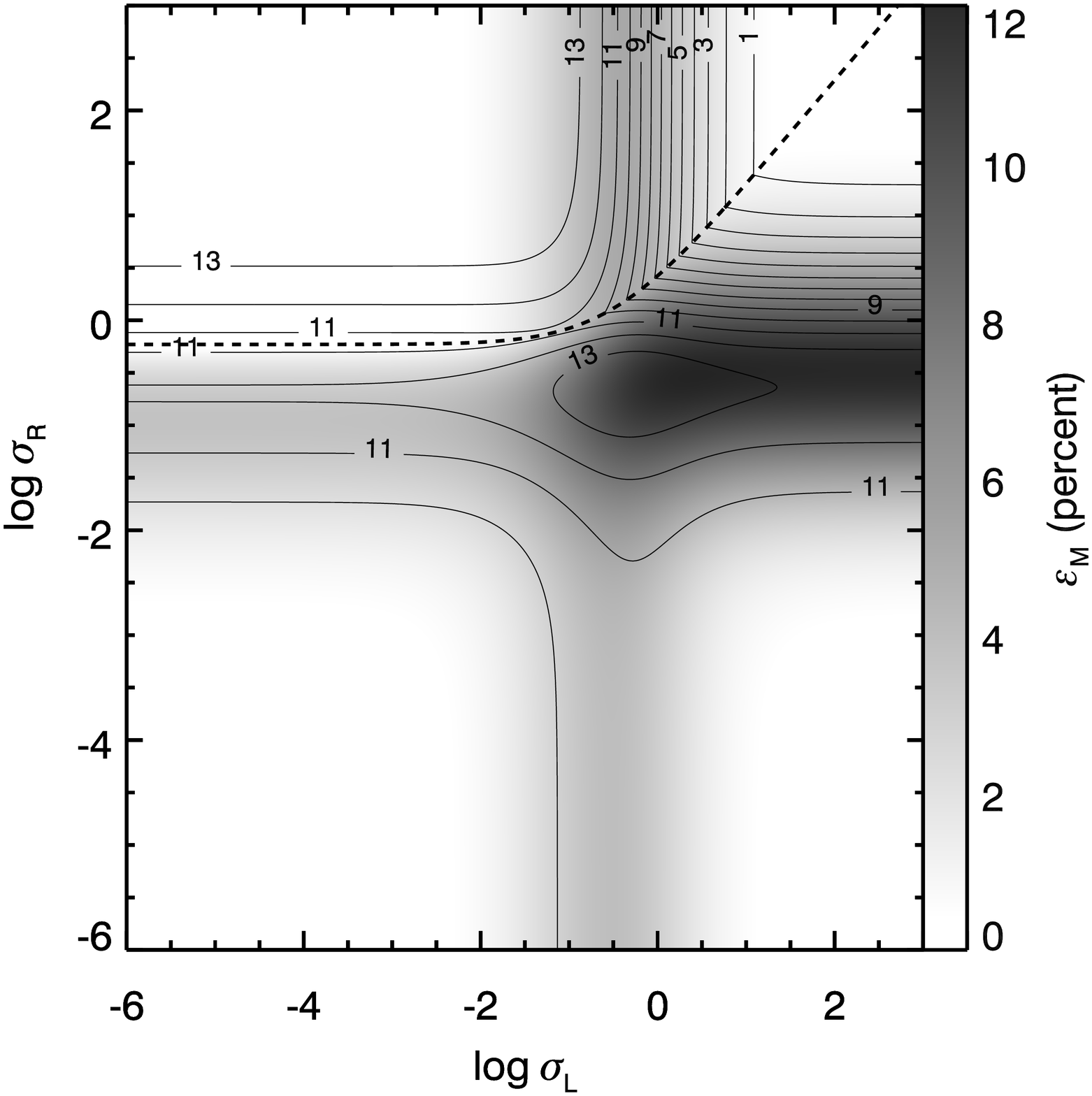}\hspace{0.2cm}
\includegraphics[width=8.5cm]{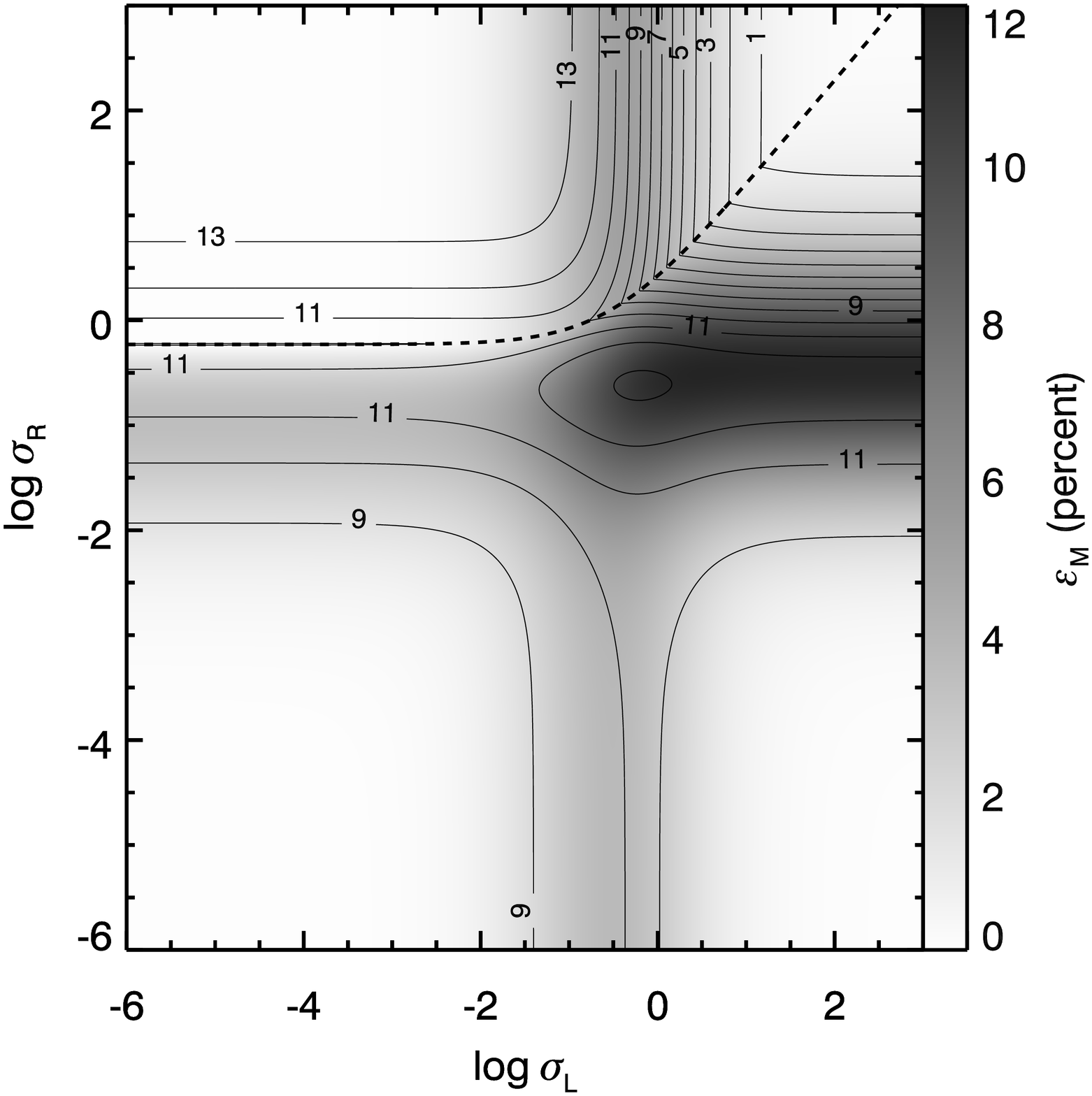}
\caption{Left panel: contours of the total dynamic efficiency for the model
  parameters from Table 1 computed using definition of MA10 (see
  Eq.~\ref{eq:energiesMA10}. Contours indicate the efficiency in
  percent and their levels are $1$, $2$, $3$, $4$, $5$, $6$, $7$, $8$,
  $9$, $10$, $11$, $12$ and $13$. In the region of the parameter space
  above the dashed line there is no forward shock. Filled contours show
  the magnetic efficiency. Right panel: same as left panel, but using
  Eq.~\ref{eq:energiesNKT11}, inspired by NKT11.}
\label{fig:dyneff}
\end{figure*}

With these definitions, the ratio of energies shown in
Eq.~\ref{eq:thermalL} becomes Lorentz invariant, and so is the total
dynamic efficiency. Furthermore, when calculated in the LAB-frame, the
exact values of the dynamical efficiency computed using any of the two
sets of definitions (Eqs.~\ref{eq:energiesMA10} or
\ref{eq:energiesNKT11}) differs very little, as it is shown in
Fig.~\ref{fig:dyneff}.

\section{Computational requirements for realistic calculations}
\label{sec:dis_comp}

\corr{In this paper we have adopted a cylindrical geometry and ignored
  adiabatic losses of the non-thermal electrons, as well as the
  high-latitude emission. However, we have included both of these
  effects in previous works studying afterglows of gamma-ray bursts
  \citep{Mimica:2010cc,Mimica:2011az}. In those calculations,
  performed using the radiative transfer code \emph{SPEV}
  \citep{Mimica:2009aa}, we have only taken into account the
  synchrotron and EIC emission processes. Even so, the calculation of
  a realistic multi-wavelength light curve of a \emph{single} model
  lasts anywhere between few hours to few days on a supercomputing
  cluster with several hundreds of computing cores, depending on the
  number of wavelengths at which the emission is computed. If the
  synchrotron-self absorption (SSA) is included, than the parallel
  scalability of the method is reduced and the calculations can become
  prohibitively expensive (see footnote 3 of \citealt{Mimica:2010cc}).

  In this paper we have a similar problem due to the fact that we
  include the SSC process, which, as SSA, is non-local and difficult
  to parallelize in a realistic conical geometry. The use of
  cylindrical geometry and other simplifying assumptions has enabled us
  to nevertheless compute light curves of $900$ models for $96$
  frequencies at $120$ observer times. This required $200$ thousand
  computing hours on the \emph{MareNostrum} computer of the Barcelona
  Supercomputing Center. Based on our previous experience, we estimate
  an order of magnitude more resources are needed if the adiabatic
  cooling of the electrons is included, and another order of magnitude
  if an accurate radiative transfer method such as \emph{SPEV} is used
  instead of approximate method described in
  Section~\ref{sec:radiation}. We note that such a calculation
  requires resources in the range of $10$ million computing hours on a
  computer such as \emph{MareNostrum}.}

\bibliographystyle{mn2e}
\bibliography{magradint.bib}

\end{document}